\documentclass[11pt]{article}
\pdfoutput=1 

\usepackage{multirow} 
\usepackage{cancel,slashed}
\usepackage{bbm}
\usepackage{amssymb}
\usepackage{amsfonts}
\usepackage{amsmath}
\usepackage{graphicx}
\usepackage{cite}
\usepackage{bbold}
\usepackage{latexsym}
\usepackage{color}
\usepackage{xypic}
\usepackage{cancel,slashed}
\usepackage[linktocpage]{hyperref}
\usepackage{color}
\usepackage{transparent}
\usepackage{footmisc}
\usepackage{mathtools}
\usepackage{array}
\usepackage{makecell}
\usepackage[utf8]{inputenc}
\usepackage{amssymb,graphics,psfrag}
\usepackage{array,epsfig,multirow,graphicx,mathtools}
\usepackage{subfigure}

\makeatletter
\def\@xfootnote[#1]{%
  \protected@xdef\@thefnmark{#1}%
  \@footnotemark\@footnotetext}
\makeatother

\newcommand{\nn}{\nonumber}

\newcommand{\bmat}{\left(\begin{array}}
\newcommand{\emat}{\end{array}\right)}

\def\d{\delta}

\def\-{\hphantom{-}}

\def\s2{\frac{1}{\sqrt2}}

\def\beq{\begin{equation}}
\def\eeq{\end{equation}}
\def\beqa{\begin{eqnarray}}
\def\eeqa{\end{eqnarray}}

\def\Tr{{\rm Tr \,}}

\def\Dsl{\,\raise.15ex\hbox{/}\mkern-13.5mu D} 

\def\CN {{\cal N}}

\def\be{\begin{equation}}
\def\ee{\end{equation}}
\def\bea{\begin{eqnarray}}
\def\eea{\end{eqnarray}}

\def\cN{\mathcal{N}}

\def\d{{\delta}}

\def\sm2{{\mbox{\small 2}}}

\newcommand{\bp}{\begin{pmatrix*}[r]}  
\newcommand{\ep}{\end{pmatrix*}}  
\newcommand{\bpp}{\begin{pmatrix}}  
\newcommand{\epp}{\end{pmatrix}}  
\newcommand{\bcd}{\begin{center}
\begin{tikzcd}}
\newcommand{\ecd}{\end{tikzcd} \end{center}}

\def\cO{\mathcal{O}}
\def\cS{\mathcal{S}}

\def\1{\mathbb{1}}

\def\d{{\rm d}}

\usepackage{tikz}
\usetikzlibrary{arrows,snakes,backgrounds}
\usetikzlibrary{shapes.misc}
\usetikzlibrary{decorations.pathreplacing}

\tikzstyle{gauge}=[circle,draw=blue!50,fill=blue!20,thick, inner sep=0pt,minimum size=1cm]
\tikzstyle{circ}=[circle,draw,thick,
inner sep=0pt,minimum size=1cm]
\tikzstyle{Dfive}=[circle, cross, draw=black!50,thick, inner sep=0pt,minimum size=0.4cm]
\tikzstyle{DfiveBig}=[circle, cross, draw=black!50,thick, inner sep=0pt,minimum size=0.6cm]
\tikzstyle{lpre}=[-,shorten <=0pt,>=stealth', very thick]
\tikzstyle{lpost}=[-,shorten >=0pt,>=stealth',very thick]
\tikzstyle{global}=[rectangle,draw=black!50,fill=black!20,thick,
inner sep=0pt,minimum size=1cm]
\tikzstyle{pre}=[<-,shorten <=0pt,>=stealth', very thick]
\tikzstyle{post}=[->,shorten >=0pt,>=stealth',very thick]
\tikzstyle{bpost}=[->, shorten >=2pt, shorten <=2pt, >=stealth', very thick]
\tikzstyle{bpre}=[<-, shorten >=2pt, shorten <=2pt, >=stealth', very thick]
\tikzstyle{nodo}=[circle,draw=black,fill=black,thick, inner sep=0pt,minimum size=2mm]
\tikzstyle{nodoblu}=[circle,draw=blue,fill=blue,thick, inner sep=0pt,minimum size=2mm]
\tikzset{cross/.style={path picture={ 
			\draw[black]
			(path picture bounding box.south east) -- (path picture bounding box.north west) (path picture bounding box.south west) -- (path picture bounding box.north east);
		}}}
		
		\tikzset{
			partial ellipse/.style args={#1:#2:#3}{
				insert path={+ (#1:#3) arc (#1:#2:#3)}
			}
		}

\begin{document}
\pagestyle{plain}

\makeatletter
\@addtoreset{equation}{section}
\makeatother
\renewcommand{\theequation}{\thesection.\arabic{equation}}
\pagestyle{empty}
\rightline{DESY 19-177}
\rightline{ROM2F/2019/06}
\vspace{0.5cm}
\begin{center}
\Huge{{The Geometry of SUSY Enhancement}
\\[15mm]}
\normalsize{Federico Carta,$^1$ Simone Giacomelli,$^{2}$ Hirotaka Hayashi,$^{3}$ and Raffaele Savelli$^4$ \\[10mm]}
\small{
${}^1$ Deutches Electronen-Synchrotron, DESY, Notkestra$\beta$e 85,\\ 22607 Hamburg, Germany. \\[2mm]
${}^2$ Mathematical Institute, University of Oxford, Andrew Wiles Building\\ Radcliffe Observatory Quarter (550), Woodstock Road,\\ Oxford, OX2 6GG, United Kingdom\\[2mm]
${}^3$ Department of Physics, School of Science, Tokai University, \\ 4-1-1 Kitakaname, Hiratsuka-shi, Kanagawa 259-1292, Japan \\[2mm] 
${}^4$ Dipartimento di Fisica, Universit\`a di Roma ``Tor Vergata'' \& INFN - Sezione di Roma2 \\ Via della Ricerca Scientifica, I-00133 Roma, Italy
\\[10mm]} 
\normalsize{\bf Abstract} \\[8mm]
\end{center}
\begin{center}
\begin{minipage}[h]{15.0cm} 

We provide a precise geometric picture that demystifies the phenomenon of supersymmetry enhancement along certain RG flows of four-dimensional field theories, recently discovered by Maruyoshi and Song. It applies to theories of arbitrary rank and it is based on a hyperk\"ahler-structure restoration on the moduli space of solutions of (twisted) Hitchin systems, which underly the class-$\cS$ construction we use as an engineering tool. Along the way, we formulate a necessary algebraic condition for supersymmetry enhancement, and, when enhancement occurs, we are able to derive the Seiberg-Witten geometry and all conformal dimensions of Coulomb-branch operators for the infrared theory, without using a-maximization.

\end{minipage}
\end{center}

\newpage
\setcounter{page}{1}
\pagestyle{plain}
\renewcommand{\thefootnote}{\arabic{footnote}}
\setcounter{footnote}{0}


\tableofcontents


\section{Introduction}
\label{s:intro}

In the last few years the phenomenon of infrared supersymmetry enhancement in quantum field theories, first observed by Maruyoshi and Song \cite{Maruyoshi:2016tqk}, has been a subject of intense investigation. 
A remarkable outcome of these studies is the discovery of four-dimensional (4d) $\mathcal{N}=1$ Lagrangian theories which flow in the infrared (IR) to non-Lagrangian $\mathcal{N}=2$ theories, often of generalized Argyres-Douglas type \cite{Argyres:1995jj, Argyres:1995xn, Eguchi:1996vu, Eguchi:1996ds, Shapere:1999xr, Cecotti:2010fi, Cecotti:2011rv, Bonelli:2011aa, Xie:2012hs, Argyres:2012fu}. Such renormalization group (RG) flows in 4d have been further generalized in \cite{Maruyoshi:2016aim, Agarwal:2016pjo, Agarwal:2017roi, Benvenuti:2017bpg, Maruyoshi:2018nod, Buican:2018ddk}, and understood more deeply in \cite{Giacomelli:2017ckh,Giacomelli:2018ziv}. Apart from being interesting in their own right, flows of this type have been used in order to compute RG protected quantities of the IR Argyres-Douglas theories, such as the superconformal index (see for example \cite{Maruyoshi:2016tqk, Maruyoshi:2016aim, Agarwal:2016pjo}).

In \cite{Carta:2018qke} three of the authors have initiated a study of the geometry underlying supersymmetry enhancement. The aim was to make manifest the deep origin of this phenomenon, which remained obscure at the field-theoretic level, and to shed light on those features a theory needs to have in order to exhibit enhancement. The main focus was on 4d rank-1 theories, which were engineered by a D3-brane probe of singular geometries in F-theory \cite{Banks:1996nj}. In this case the Seiberg-Witten (SW) curve \cite{Seiberg:1994rs, Seiberg:1994aj} of the field theory on the D3-brane may be identified with the elliptic fibration of the F-theory geometry. While higher-rank theories may also be realized in this context by simply adding more D3 probes, one inevitably looses the identification of the elliptic fiber of the F-theory space with the SW curve of the field theory, which was a key aspect of the construction proposed in \cite{Carta:2018qke}. The principal goal of the present paper is to make instead use of the class-$\mathcal{S}$ realization of 4d $\cN=2$ field theories \cite{Gaiotto:2009we,Gaiotto:2009hg, Cecotti:2011rv, Bonelli:2011aa, Xie:2012hs} in order to generalize the geometric investigation of  \cite{Carta:2018qke} to higher-rank theories. 

Let us first briefly recall the Maruyoshi-Song procedure. One usually starts with a (not necessarily Lagrangian) $\cN=2$ theory in 4d, and deforms it by adding the superpotential coupling 
\be\label{MSdef}
\delta W=\Tr(\mu M)\,,
\ee
where $M$ is a chiral multiplet in the adjoint representation of the flavor group we add by hand, $\mu$ is the moment map of the flavor symmetry, and the trace is evaluated over flavor indices. The field $M$ is then given a nilpotent vacuum expectation value (vev) and the deformation halves the amount of preserved supersymmetry. Nevertheless, for specific choices of initial theory and nilpotent vev, such a deformation triggers an RG flow which leads (upon getting rid of a bunch of free fields) to a new (typically non-Lagrangian) $\cN=2$ theory in the IR.

One general conclusion that may be inferred from the analysis of \cite{Carta:2018qke} is that the phenomenon of supersymmetry enhancement in the IR (at least the one originating from the above-sketched procedure) seems to be intimately correlated with the local structure of some auxiliary algebraic space $X_n$ around the origin. More precisely, if supersymmetry is to enhance, some non-trivial factorization needs to take place, which turns said space locally into a product of a lower-dimensional space times a trivial factor, i.e.~$X_n\simeq Y_{n-p}\times\mathbb{C}^p$.\footnote{In all interesting cases the origin carries some singularity, because any space trivially factorizes in the neighborhood of a smooth point.} What the space $Y_{n-p}$ exactly is and what the factorization of $X_n$ precisely means depends on the context. For instance, in the case of field theories in three space-time dimensions, $Y_{n-p}$ is identified with the Coulomb branch (CB) of the moduli space, which gets geometrized in M-theory. The enhancement phenomenon is then directly explained in purely geometric terms as a \emph{holonomy reduction} of $X_n$, which is the space probed by M2-branes in M-theory.

In the 4d case, however, $X_n$ and $Y_{n-p}$ are rather auxiliary spaces, whose interpretation changes according to the way one geometrically engineers the field theory. For example, if one uses F-theory to realize rank-1 theories (as was the case in \cite{Carta:2018qke}), $X_n$ is nothing but the elliptically-fibered geometry probed by a D3-brane, and again supersymmetry enhances only when this space locally exhibits a holonomy reduction (in this case down to $SU(2)$). In contrast, if one considers theories of class-$\mathcal{S}$ of arbitrary rank (as we are going to do here), the meaning of these spaces is more subtle. $Y_{n-p}$ is closely related to the moduli space of solutions of the Hitchin system underlying the class-$\mathcal{S}$ construction. More precisely, as is well known, for a rank-$r$ theory such a moduli space has the structure of a $T^{2r}$ fibered over the $r$-dimensional CB of the four-dimensional theory (Hitchin fibration) \cite{Kapustin:1998xn, Gaiotto:2009hg}. Associated to it there is another fibration (now $r+1$-dimensional) over the same base, whose generic fiber is the spectral curve of the Hitchin field: The genus of this curve turns out to be precisely $r$, and this nicely encodes the SW geometry of the 4d theory. In analogy with the F-theory case, we identify $Y_{n-p}$ with the latter fibration. A necessary condition for supersymmetry to enhance is then that the space $X_n$ (which also has the structure of a Riemann-surface fibration) locally factorizes in such a way that the genus of the fiber of $Y_{n-p}$ coincides with the dimension of its base. Only if this non-trivial factorization takes place, can we associate to $Y_{n-p}$ a Hitchin fibration, which will determine the low-energy dynamics  of the ensuing $\cN=2$ theory in the IR.

The above reasoning suggests that, for class-$\mathcal{S}$ theories, we can geometrically explain the enhancement in terms of the \emph{restoration} of a \emph{hyperk\"ahler structure} on the moduli space of solutions of the corresponding Hitchin system. As we will explain in more detail in Section \ref{s:Hitchin}, the starting 4d $\cN=2$ theory will be associated to a Hitchin system on a two-sphere with one regular and one irregular punctures, whose Hitchin field $\Phi$ is a meromorphic section of $\cO(-2)$ encoding the CB operators in (some of) its Casimir invariants \cite{Wang:2015mra,Giacomelli:2017ckh}. The deformation \eqref{MSdef} turns said Hitchin system into a \emph{generalized} one \cite{Xie:2013gma,Bonelli:2013pva, Xie:2013rsa, Yonekura:2013mya}, consisting of two Hitchin fields $\Phi_1,\Phi_2$ which are now meromorphic sections of $\cO(-1)$, each being singular at just one of the two punctures. In particular, in the neighborhood of the regular puncture where $\Phi_2$ has a pole, $\Phi_1$ plays the role of the field $M$. In this context, a deformation leading to supersymmetry enhancement corresponds to giving $\Phi_1$ a nilpotent vev along the \emph{principal} orbit, which forces $\Phi_2$ to become a holomorphic section of $\cO(-1)$ on the two-sphere, and thus to vanish identically. We are therefore left with a \emph{twisted} Hitchin system, whose solutions with given boundary conditions at the irregular puncture\footnote{We restrict our attention to irregular punctures with the property that boundary conditions for the Hitchin field can be univocally inferred from its characteristic polynomial. Our approach is not refined enough to treat irregular punctures where the Hitchin field has degenerate eigenvalues, i.e.~the so-called type III punctures \cite{Xie:2012hs}. We will briefly comment about them in Section \ref{s:Conclusions}.}, as we will show, are in bijection with those of an ordinary Hitchin system. This strongly suggests that a hyperk\"ahler structure can be restored on the corresponding moduli space, hence explaining why supersymmetry enhances.

Armed with this understanding of the geometry underlying supersymmetry enhancement, in Section \ref{s:Systematics} we will carry out a systematic analysis of the Maruyoshi-Song flows. Given the starting theory and the nilpotent orbit, the ``interpolating'' geometry $X_n$ remains the same at all energy scales, and the IR behavior of the theory crucially depends on a possible factorization $X_n\simeq Y_{n-p}\times\mathbb{C}^p$. Along the lines of \cite{Carta:2018qke}, on the one hand we will derive a simple algebraic criterion to rule out supersymmetry enhancement. On the other hand, for the cases that exhibit supersymmetry enhancement we will derive, using purely algebraic methods, the correct scaling dimensions of CB operators as well as the explicit form of the infrared SW geometry, including all masses and couplings.

The paper is organized as follows: In Section \ref{s:Hitchin}, after reviewing some material about $\cN=1$ class-$\mathcal{S}$ theories, we demonstrate how, for flows exhibiting SUSY enhancement, the generalized Hitchin system turns into a twisted Hitchin system in the IR. Its solutions are shown (Subsec. \ref{ss:bijection}) to be in bijection with those of an ordinary system. In Section \ref{s:Systematics} we derive in a purely algebraic manner two necessary criteria for enhancement. Rather than stating them abstractly, we present them in the context of two specific Lagrangian models, in particular $\cN=2$ SQCD with gauge group $SU(3)$. Finally, we draw our conclusions in Section \ref{s:Conclusions} and briefly comment on open issues related to punctures of type III.

\section{SUSY enhancement and Hitchin systems}
\label{s:Hitchin}

In this section we explain how, in the context of $\cN=1$ class-$\mathcal{S}$ field theories, supersymmetry enhancement originates from the emergence of an ordinary Hitchin system out of a generalized Hitchin system. The key intermediate step will be a bijection between solutions to the ordinary Hitchin system and solutions to a suitably twisted one. Our method uses deformations induced by principal nilpotent vevs only. Since there are cases of non-principal deformations leading to enhancement too, in Subsection \ref{ss:non-principal} we will explain how our approach allows us to recover those.

\subsection{Preliminaries}

Let us consider M-theory on the background $\mathbb{R}^4\times X\times\mathbb{R}$, where $X$ is a Calabi-Yau threefold. 
A stack of $N$ M5 branes wrapping $\mathbb{R}^4\times\mathcal{C}$, where $\mathcal{C}$ is a holomorphic two-cycle in $X$, describes an $\cN=1$ theory on 
$\mathbb{R}^4$. We consider backgrounds of the form $$X=L_1\oplus L_2\,,$$ 
where $L_1$ and $L_2$ are holomorphic line bundles on $\mathcal{C}$ of degree $p$ and $q$. Indeed the Calabi-Yau 
condition imposes the constraint $p+q=2g-2$, where $g$ is the genus of the Riemann surface. In this paper we will be concerned only with theories for which $\mathcal{C}$ is a sphere and therefore, from now on, we will restrict to this case. The two line bundles then satisfy the constraint $p+q=-2$, reflecting the Calabi-Yau condition
\begin{equation}\label{tensorprd}
L_1\otimes L_2= \mathcal{O}(-2)\,.
\end{equation}
As in the $\cN=2$ case, a sphere with an arbitrary number of regular punctures can be thought of as a collection of trinions (spheres with three punctures) connected together, where connecting two trinions together is physically interpreted as gauging the diagonal combination of their global symmetry. The gauging can be either $\cN=1$ or $\cN=2$ depending on the details of the geometric construction \cite{Benini:2009mz, Bah:2011je, Bah:2011vv, Bah:2012dg, Fazzi:2016eec, Nardoni:2016ffl}. In order to describe the resulting four-dimensional theory, it therefore suffices to understand what the trinions are. 

In the special case of trinions with punctures which (locally) preserve 8 supercharges\footnote{These are the punctures appearing in the standard $\cN=2$ Class $\mathcal{S}$ construction and correspond to the $1/2$ BPS boundary conditions for $\cN=4$ SYM \cite{Gaiotto:2008ak}. In principle one could consider more general punctures corresponding to $1/4$ BPS boundary conditions \cite{Hashimoto:2014vpa, Hashimoto:2014nwa}, but we will not need this in our paper.} we can proceed as follows: We decompose a trinion into a sphere with three holes (pair of pants) and three caps with a puncture. For each of these building blocks we take the canonical and the trivial line bundles. When we connect a cap to the pair of pants, we also need to specify how the corresponding line bundles are glued together: We can either glue the canonical bundles (and therefore the trivial bundles) together, or we can glue the canonical bundle of one block to the trivial bundle of the other. Once we have done that, we end up with our trinion endowed with the two line bundles $L_{1,2}$.

We can encode these geometric data by attaching a sign to each puncture and to the pair of pants. When the signs of the puncture and of the pair of pants agree, it means that we are gluing the corresponding canonical bundles together. Of course, if we change the sign of all the building blocks we are simply interchanging $L_1$ and $L_2$ and we end up with the same theory.
 We easily see that if all building blocks are of the same kind, one line bundle gets identified with the trivial bundle on the 
sphere (and the other with its canonical bundle) and the threefold is of the form $T^*(S^2)\times\mathbb{C}$. This special case corresponds to an $\cN=2$ class $\mathcal{S}$ trinion.  

A nice feature of this construction is that the degrees of the two line bundles $L_{1,2}$ can be computed straightforwardly: The first Chern class receives a nontrivial contribution only from the canonical bundle on the various building blocks and we simply need to sum the various contributions. The canonical bundle of a cap contributes $-1$ whereas the sphere with three holes contributes $+1$. In any case the constraint $p+q=-2$ is always automatically satisfied. 

Let us consider $T_N$ theory, which has three full punctures and all the building blocks have the same sign. If we now modify the theory by changing the sign of one puncture the two line bundles become $L_1\equiv L_2=\mathcal{O}(-1)$. Physically, this is interpreted as follows: We start from $T_N$ and we add by hand a chiral multiplet $M$ transforming in the adjoint representation of the global symmetry carried by the puncture. We also couple it to the corresponding moment map $\mu$ by adding the superpotential term $\Tr(\mu M)$. Indeed we can generalize the construction by including generic punctures, which are in one-to-one correspondence with nilpotent orbits of the global symmetry. When the signs of the pair of pants and the puncture agree and the puncture is not full, it means that we have higgsed the theory with a full puncture by turning on a nilpotent vev for the corresponding moment map. If instead the signs do not agree, it means that we have turned on a nilpotent vev for the singlet $M$ rather than the moment map, which is now set to zero in the chiral ring due to the F-term equation for $M$. Combining these operations we can construct all of the $\cN=1$ trinions starting from $T_N$ plus a collection of chiral multiplets. 

The moduli space of these $\cN=1$ theories (on $\mathbb{R}^3\times S^1$) is described by the solutions of a generalized Hitchin system involving two Hitchin fields ($\Phi_1$ and $\Phi_2$) which are sections of the line bundles $L_1$ and $L_2$ respectively. The equations of the generalized Hitchin system state that these fields are covariantly holomorphic and commute ($[\Phi_1,\Phi_2]=0$). Each field is singular at punctures of a given sign only (for example $\Phi_1$ is singular only at punctures with sign plus and analogously $\Phi_2$ is singular only at punctures with sign minus). The singularity is the same as in the $\cN=2$ case. Indeed, in the $\cN=2$ case, one field is a one-form and is singular at all the punctures, whereas the other is a function without poles and is therefore constant. Setting it to zero we recover the description of the Coulomb branch of the $\cN=2$ theory in terms of an ordinary Hitchin system.

In the rest of this paper we will be concerned with $D_k^b(J)$ theories, which correspond to a sphere with two punctures, one is full and the other is irregular \cite{Wang:2015mra,Giacomelli:2017ckh}. $J$ is an ADE group and labels the choice of the six-dimensional $\cN=(2,0)$ theory we compactify on the sphere. The parameters $k$ and $b$ specify the choice of irregular puncture: If we take a local coordinate $w$ on the sphere such that the irregular puncture is located at $w=0$, the behavior of the Hitchin field near $w=0$ is 
\be\Phi\simeq \frac{T}{w^{1+k/b}}+\dots\,,\ee 
where $T$ is a  regular semi-simple element of the Lie algebra $J$ and the dots stand for less singular terms.\footnote{Note that our notation slightly differs from the one adopted in \cite{Wang:2015mra}, whereby the parameter $k$ is shifted by one unit of $b$ with respect to the $k$ appearing here.} The parameter $k$ is an arbitrary positive integer, whereas $b$ can take two or three different values depending on the choice of $J$:
\be\label{sing}\begin{array}{|c|c|}
\hline
J & b \\
\hline 
A_{N-1} & N;\; N-1 \\
\hline
D_N & 2N-2;\; N \\
\hline 
E_6 & 12;\; 9;\; 8 \\
\hline 
E_7 & 18;\; 14 \\
\hline 
E_8 & 30;\; 24;\; 20 \\
\hline
\end{array}\ee
Notice that the Coxeter number $h(J)$ is always an allowed value for $b$. In the following we will drop the label $b$ whenever $b=h(J)$. A detailed discussion about these theories can be found in \cite{Giacomelli:2017ckh}.

The analysis with the two Hitchin fields briefly reviewed above does not immediately apply to $D_k^b(J)$ models. However, by analogy with the case of $\cN=1$ theories labelled by a sphere with regular punctures only, we propose that the $D_k^b(J)$ theory deformed by coupling an adjoint chiral to the moment map associated with the symmetry carried by the full puncture is described by a generalized Hitchin system in which both fields are sections of $\mathcal{O}(-1)$. One field is singular at the irregular puncture only (say $\Phi_1$) and the singularity is the same as in the parent $\cN=2$ theory, whereas the other field $\Phi_2$ is singular at the regular puncture only. Moreover, giving a nilpotent vev to the adjoint chiral (i.e.~initiating a Maruyoshi-Song flow) can be implemented by changing the boundary condition for $\Phi_2$. The nontrivial consistency checks we will find below give strong evidence in favor of our claim.

\subsection{RG flows and spectral curves}

\subsubsection*{Extracting the SW curve of the IR theory}

In this section we will use the results reviewed in the previous section about the generalized Hitchin system to analyze the  Maruyoshi-Song flow at the level of the SW curve. As is well known, in the case of the ordinary Hitchin system the SW curve for the underlying $\CN=2$ theory is encoded in the spectral equation for the Hitchin field $\Phi$ \cite{Gaiotto:2009hg}: 
\be\label{specc} \det(\lambda-\Phi)=0\,,\ee 
where $\lambda$ is the SW differential. If we now choose local coordinates for the base and fiber of $T^*(\mathcal{C})$ and write $\lambda$ in terms of those, (\ref{specc}) becomes the SW curve describing the theory. 
In the case of $\CN=1$ class $\mathcal{S}$ theories we have a similar result involving the spectral equations of the generalized Hitchin fields $\Phi_{1,2}$ \cite{Xie:2013gma,Bonelli:2013pva, Xie:2013rsa, Yonekura:2013mya}: 
\be\label{genhit}\left\{\begin{array}{l} 
  \det(\lambda_1-\Phi_1)=0  \\
  \det(\lambda_2-\Phi_2)=0\,, \\
\end{array}\right. \ee 
where $\lambda_{1,2}$ are sections of the corresponding line bundles. In general the system (\ref{genhit}) should then be supplemented by further equations enforcing the commutativity constraint $[\Phi_1,\Phi_2]=0$. As we will explain later, this fact will not play any role in our discussion.

Now we use our guess that the $D_k^{b}(J)$ theory with a chiral multiplet in the adjoint of $J$ coupled to the corresponding moment map is described by a generalized Hitchin system, with both line bundles of degree $-1$. One field, say $\Phi_1$, is singular at the irregular puncture only, whereas $\Phi_2$ is singular at the regular puncture. The boundary conditions at the two punctures are the same as in the parent $\CN=2$ theory. Upon giving a principal nilpotent vev to the adjoint chiral we remove the regular puncture completely. As a result, in the geometry describing the infrared fixed point, the field $\Phi_2$ becomes a section of $\mathcal{O}(-1)$ on the sphere without poles and therefore vanishes identically. We can therefore focus on the spectral equation of $\Phi_1$ only. 

Let us illustrate the procedure for the class of theories $D_k(SU(N))$ (i.e. $J=SU(N)$ and $b=N$). The extension to other models with $J=SU(N)$ or $J=SO(2N)$ is trivial. The SW curve and differential read 
$$x^N+z^k+...=0\,;\qquad \lambda=x\frac{dz}{z}\,,$$ 
where the dots stand for subleading terms. We can rewrite it as in (\ref{specc}): 
\be\lambda^N+\sum_{\alpha=2}^N\lambda^{N-\alpha}P_{d_{\alpha}}(z)\left(\frac{dz}{z}\right)^{\alpha}=0\,.\ee 
The polynomials $P_{d_{\alpha}}(z)$ have degree $d_{\alpha}$ equal to the integer part of $k\alpha/N$ and $P_{d_N}(z)$ can be taken to be monic. The degree $\alpha$ differentials have a pole of order $\alpha$ at $z=0$ (the regular puncture) and a pole of order $\alpha+d_{\alpha}$ at $z=\infty$ (irregular puncture). In order to model the infrared fixed point of the Maruyoshi-Song flow we now turn our attention to a twisted Hitchin field which is a section of $\mathcal{O}(-1)$ and is singular only at infinity, where the irregular puncture is located. The corresponding spectral equation then reads 
\be\label{twhit}\lambda_1^N+\sum_{\alpha=2}^N\lambda_1^{N-\alpha}P_{d_{\alpha}}(z)(dz)^{\frac{\alpha}{2}}=0\,,\ee 
where the various terms are chosen to reproduce the singular behaviour at the irregular puncture\footnote{We denote with $(dz)^{\alpha/2}$ a section of $\mathcal{O}(-\alpha)$ on the sphere without zeros and with a pole of order $\alpha$ at infinity.}. 

Our claim now is that the twisted Hitchin field $\Phi_1$ whose spectral equation is given by (\ref{twhit}) is equivalent to an ordinary Hitchin field $\widetilde{\Phi}$ obtained by tensoring $\Phi_1$ with a reference section of $\mathcal{O}(-1)$ having a simple pole at infinity. The corresponding spectral equation is then obtained by tensoring (\ref{twhit}) with $(dz)^{N/2}$: 
\be\label{newhit}\widetilde{\lambda}^N+\sum_{\alpha=2}^N\widetilde{\lambda}^{N-\alpha}P_{d_{\alpha}}(z)(dz)^{\alpha}=0\,.\ee 
This equation precisely encodes the SW data of the IR fixed point of the RG flow, namely the theory $(A_{N-1},A_{k-1})$. In order to see this, we choose local coordinates on $T^*(\mathbb{P}^1)$ and set $\widetilde{\lambda}=\tilde{x}dz$. Plugging this into (\ref{newhit}) we find 
\be\tilde{x}^N+z^{k}+...=0\,;\quad \widetilde{\lambda}=\tilde{x}dz\,.\ee 
 We normalized the coordinates in such a way that $P_{d_N}(z)$ in (\ref{newhit}) is monic. We can also take advantage of the freedom to shift $z$ (which does not change the SW differential up to exact terms) to remove all subleading terms proportional to $z^{k-1}$. These are precisely the SW curve and differential of the $(A_{N-1},A_{k-1})$ theory.

\subsubsection*{Counting decoupled operators} 

In order to count decoupled operators we can make use of the one-to-one correspondence between UV and IR CB operators discussed in \cite{Giacomelli:2018ziv}, which we will now review.
We start by recalling that for $D_k(SU(N))$ theories the versal deformations of the $A_{N-1}$ singularity are the mass Casimirs of the $SU(N)$ global symmetry. The vev of ultraviolet (UV) CB operators is instead described by the $z$-dependent deformation terms. The $(A_{N-1},A_{k-1})$ theory is described by the same curve but the CB operators correspond to all deformation terms with coefficient of dimension larger than one. The one-to-one correspondence between UV and IR CB operators is then described as follows: given any UV CB operator $u$, divide the corresponding deformation term by $z$. This operation maps the original term to another deformation and the scaling dimension of the corresponding parameter $u'$ is that of $u$ plus the dimension of $z$, which in the $D_k(SU(N))$ theory is equal to $\frac{N}{k}$. Since by assumption $D(u)>1$, we conclude that $$D(u')>1+\frac{N}{k}=\frac{k+N}{k}\,.$$ 
Now we exploit the observation that the scaling dimension in the IR of $u'$ (provided it does not decouple) is $D(u')$ times $\frac{k}{k+N}$,\footnote{This can be seen e.g. by comparing the deformation term of highest dimension in the UV and in the IR. See \cite{Giacomelli:2017ckh} for details.} and due to the above inequality, we clearly see that this quantity is larger than one. We then conclude that the term $u'$ always corresponds to a CB operator of the IR theory $(A_{N-1},A_{k-1})$. Analogously, coupling constants of dimension smaller than one in the UV are mapped to coupling constants in the IR and mass parameters of dimension one are mapped to mass parameters. 

Due to the fact that the curve describing UV and IR fixed points are the same, the deformation parameters in the two cases are clearly equal in number. In the UV there are $N-1$ parameters (the mass Casimirs of the $SU(N)$ global symmetry, which in the $\mathcal{N}=1$ theory are rather interpreted as expectation values for the singlets) on which the map described above is not defined, and accordingly we expect to see $N-1$ parameters in the IR which do not arise from our UV-IR map. These are easy to describe: For any integer $n\leq N-2$ find the largest $j$ such that the monomial $x^nz^j$ appears in the curve\footnote{For $n=0$ we take $j$ to be $k-1$.}. Clearly all such terms (and only those) cannot arise from our map and they are precisely $N-1$ in number. Terms of the form $x^nz^{k-1}$ can actually be removed by shifting $z$\footnote{This change of variables is allowed in the IR only because in the UV it would change the location of the regular singularity.} and do not arise in the infrared theory. Their number can be easily determined to be the integer part of $N/k$ plus one. All the other terms correspond to coupling constants in the infrared theory. This is seen as follows: the dimension of the parameters multiplying the monomials $x^nz^j$ has to be smaller than the dimension of $z$, otherwise $x^nz^{j+1}$ would be an allowed deformation term. Combining this with the fact that the dimension of $z$ in the IR is smaller than 1, we reach the desired conclusion. So we conclude as expected that, out of the singlets and UV CB operators, all except $N-1$ operators become CB operators in the IR.

\subsection{Twisted vs ordinary Hitchin systems}\label{ss:bijection}

The analysis of the previous section relied on the equivalence of the moduli space of solutions of two different Hitchin systems on a punctured Riemann sphere: An ordinary one, with Hitchin field $\widetilde{\Phi}\in \Gamma (\mathcal{O}(-2))$, and a twisted one, with Hitchin field $\Phi_1\in \Gamma (\mathcal{O}(-1))$. Boundary conditions are such that both these Hitchin fields are smooth sections everywhere on the sphere except at one point (the same point for both), the irregular puncture, where they develop a pole. We argue that there is a one-to-one correspondence between solutions of the two systems with said boundary conditions. One way to see this is to bijectively map one system of equations to the other.

The non-holomorphic equation of an ordinary Hitchin system reads
\begin{eqnarray}\label{Ordinary}
F+[\widetilde{\Phi},\widetilde{\Phi}^\dagger]=0\,,
\end{eqnarray}
where $F$ is the $(1,1)$-form gauge field strength and $^\dagger$ simply denotes complex conjugation and matrix transposition. This equation can also be trivially written in terms of components and, in the local patch $U_z$ with coordinate $z$, it takes the simple form:
\be
F_{z\bar{z}}+[\widetilde{\Phi}_z,\widetilde{\Phi}^\dagger_{\bar{z}}]\,.
\ee
On the contrary, the non-holomorphic equation of a twisted Hitchin system only has a well defined expression in terms of components:
\be\label{twisted}
h^{-1/2}_{z\bar{z}}F_{z\bar{z}} +[(\Phi_1)_z,(\Phi_1)^\dagger_{\bar{z}}]\,,
\ee
where $h$ is the hermitian metric on the tangent bundle of the sphere.

Let us now write
\be\label{Isomorph}
\widetilde{\Phi}=\hat{s} \,\Phi_1\,,
\ee
where $\hat{s}\equiv s/||s||$, and $s$ is a nowhere-vanishing reference section of $\mathcal{O}(-1)$ that is smooth everywhere except at one point, which we choose to be the same point where $\widetilde{\Phi}$ and $\Phi_1$ are singular. This condition on $s$ is needed in order not to change the assigned boundary conditions of the two Hitchin fields by creating further poles. There is only one such reference section (modulo rescaling by global smooth functions), and it has obviously a pole of order $1$ at the irregular puncture. For later convenience, we normalized this section by dividing it by its norm, i.e.~by the square root of the globally well-defined smooth function
\be\label{norm}
||s||^2=h^{-1/2}\bar{s}s\,.
\ee
Placing the irregular puncture at the infinity of the patch $U_z$, the local presentation of $\hat{s}$ in that patch is
\be\label{locals}
\hat{s}|_{U_z} = \frac{\sqrt{{\rm d}z}}{||\sqrt{{\rm d}z}||}\,,\qquad\quad ||\sqrt{{\rm d}z}||^2=\sqrt{\frac{{\rm d}z\,{\rm d}\bar{z}}{1+|z|^2}}\,,
\ee
where we have used the Fubini-Study metric on $\mathbb{P}^1$ to write down the norm.

Plugging \eqref{Isomorph} into \eqref{Ordinary} and using \eqref{norm} yields \eqref{twisted}. Since $\hat{s}$ is nowhere vanishing and unique, \eqref{Isomorph} is a bijective map between solutions to the ordinary and the twisted Hitchin systems. This map respects the boundary conditions of the Hitchin fields, but it changes their order of pole at the irregular puncture. To see this, it is convenient to work in the local patch $U_w$, where $w=1/z$, so that the irregular puncture is located at $w=0$. Locally we can always switch to a gauge, the holomorphic gauge, where $A^{0,1}=0$, and thus have the Hitchin field satisfy the equation
\be
\bar{\partial}_{\bar{w}} \widetilde\Phi_w= 2\pi i \sum_{i=0}^{p-1}a_i\partial_w^i\delta_w\,,
\ee
where $\delta_w$ is the delta function on $w=0$ and $a_{i}$ are matrix-valued coefficients determining the singular behavior of $\widetilde\Phi$ such that, around the irregular puncture
\be
\widetilde\Phi \sim {\rm d}w \sum_{i=0}^{p-1} \frac{(-1)^i i!\, a_i}{w^{i+1}}\,.
\ee
Using \eqref{locals}, it is immediate to see that our reference section around the irregular puncture is
\be
\hat{s}|_{U_w}=\frac{1}{||\sqrt{{\rm d}w}||} \frac{\sqrt{{\rm d}w}}{w}\,,
\ee
and therefore we have
\be
\Phi_1 \sim \sqrt{{\rm d}w} \sum_{i=0}^{p-2} \frac{(-1)^i i!\,b_i}{w^{i+1}}\,,
\ee
where $b_i=||\sqrt{{\rm d}w}||a_{i+1}$. Hence the twisted Hitchin field has a pole at the irregular puncture of order one unit less than the one of the ordinary Hitchin field, i.e.
\be
\bar{\partial}_{\bar{w}} (\Phi_1)_w= 2\pi i \sum_{i=0}^{p-2}b_i\partial_w^i\delta_w\,.
\ee

 \subsection{Comments about non-principal nilpotent vevs}\label{ss:non-principal}
 
 In \cite{Agarwal:2016pjo} the authors found several examples of theories which exhibit supersymmetry enhancement in the IR upon turning on a non-principal nilpotent vev. At first sight these flows do not seem to fit in our discussion since under a non-principal nilpotent vev the regular puncture is not removed completely and the generalized Hitchin system does not reduce to a simpler twisted Hitchin system. The scope of this section is to notice that, as was already pointed out in \cite{Giacomelli:2017ckh}, we are not actually missing any of the known enhancing RG flows by focusing on our setup.

The argument is based on the simple observation that for any group we can get several non-principal nilpotent orbits just by embedding the principal nilpotent orbit of a subgroup. The point is the following: In our geometric setup we are activating an expectation value for the moment map associated with the full puncture only and since the global symmetry carried by the regular puncture is in general only a subgroup of the actual global symmetry of the theory, by considering the principal nilpotent vev for the corresponding moment map we are actually considering (in general) a non-principal nilpotent vev for the theory. 

Our main observation is that whenever there are multiple choices of nilpotent vev which lead to supersymmetry enhancement in the IR, there are also multiple realizations of the theory in the $D_k^b(J)$ class (with different $J$) and, by considering the principal nilpotent orbit for $J$ in the various realizations, we always recover all the enhancing RG flows. We do not have an a priori proof of this statement, but we will now check that we do recover all the RG flows discussed in \cite{Agarwal:2016pjo}.
 
\begin{itemize}
\item Let us start by the case of $SU(2)$ SQCD, which has three different realizations in the $D_k^b(J)$ class: It is equivalent to $D_1^4(SO(8))$, $D_2(SU(4))$ and $D_2^2(SU(3))$. We therefore predict that the theory exhibits enhancement upon turning on a principal nilpotent vev and also $SU(4)$ and $SU(3)$ induced nilpotent vevs respectively. It is well-known that there is a unique way to embed $SU(3)$ inside $SO(8)$ up to conjugation and the corresponding induced nilpotent orbit is labelled by the partition $[3^2,1^2]$. Indeed it was found in \cite{Agarwal:2016pjo} that the corresponding vev does lead to enhancement in the IR. The remaining enhancing orbits are the principal (as our construction correctly predicts), the orbit $[5,1^3]$ and the two $[4,4]$ orbits. The last three all lead to the same IR fixed point. This result is perfectly consistent with our construction, which predicts enhancement in the case of an $SU(4)$ induced nilpotent vev: There are three inequivalent embeddings of $SU(4)$ inside $SO(8)$ and the corresponding nilpotent orbits are precisely the three listed above. 
\item Let us now discuss the other Lagrangian cases. The only relevant ones are $SU(N)$ and $USp(2N)$ SQCD, since all other Lagrangian theories exhibit enhancement upon turning on a principal nilpotent vev only (or do not exhibit enhancement at all). In the case of $SU(N)$ SQCD ($N>2$) with $2N$ flavors there are two choices of nilpotent vevs (principal and subregular), and accordingly we have two different realizations of conformal $SU(N)$ SQCD in our class: $D_2(SU(2N))$ and $D_2^{2N-2}(SU(2N-1))$. Analogously, the two possible choices of nilpotent vev for $USp(2N)$ conformal SQCD, whose global symmetry is $SO(4N+4)$, correspond to the two different realizations of this theory in the $D_k^b(J)$ class: $D_1^{2N+2}(SO(4N+4))$ and $D_2(SO(4N+2))$. 
 \item Let us consider the theories called $(I_{N+1,1-N},F)$ in \cite{Agarwal:2016pjo}, whose global symmetry is $SU(N+1)$. These models exhibit enhancement both for the principal and subregular nilpotent orbits. When $N$ is odd the theory is equivalent to $SU(n_c)$ conformal SQCD with $n_c=\frac{N+1}{2}$ colors, whereas for $N$ even the model is not Lagrangian. We indeed recover this result! In our notation these theories have the following realizations: $D_2(SU(N+1))$ and $D_2^{N-1}(SU(N))$. The full global symmetry is manifestly visible in the first realization only. When $N=2$ the theory coincides with $D_4$ Argyres-Douglas theory (sometimes called $\mathcal{H}_2$), which has $SU(3)$ global symmetry and flows to $\mathcal{N}=2$ SCFTs under both choices of nilpotent vev (principal and minimal). 
 \item Finally, let us discuss Minahan-Nemeschansky theories. In the case of the $E_6$ theory the authors of \cite{Agarwal:2016pjo} found that there are three choices of nilpotent vev which lead to supersymmetry enhancement in the IR.
 Accordingly, it turns out that the $E_6$ Minahan-Nemeschansky theory appears three times in the $D_k^b(J)$ class: It is equivalent to $D_2(SO(8))$, $D_1^{5}(SO(10))$ and $D_1^9(E_6)$. By activating a principal nilpotent vev for the group $J$ we recover the three enhancing RG flows. We find instead two different realizations of the $E_7$ Minahan-Nemeschansky theory: $D_1^{14}(E_7)$ and $D_1^{8}(E_6)$, in agreement with the fact that enhancement occurs only for two choices of nilpotent vev. Finally, $E_8$ Minahan-Nemeschansky theory exhibits enhancement only in the case of a principal nilpotent vev. As expected we find just one realization of this model: $D_1^{24}(E_8)$.   
\end{itemize}

\section{Systematics of SUSY enhancement}
\label{s:Systematics}

In this section we will derive a necessary algebraic criterion for supersymmetry enhancement and, in case enhancement occurs, explain how to systematically derive the SW curve and differential of the IR theory (as well as the correct conformal dimensions of CB operators) without using any maximization procedure. After discussing a few general facts about the underlying geometries in Subsection \ref{subs:Prel}, we will study in detail a specific rank-$2$ Lagrangian case in Subsection \ref{subs:Su3}, in order to illustrate the key steps of our approach. We will then conclude by analyzing in Subsection \ref{ExNotEnhance} a particular rank-$6$ case, whose peculiarities will lead us to an important refinement of our algebraic criterion.

\subsection{$\mathcal{N}=1$ curves from branes}\label{subs:Prel}

For the analysis of this section it is crucial to understand how to implement the SUSY breaking deformations at the level of the underlying SW geometry. This will directly generalize the results of \cite{Carta:2018qke} for rank-1 theories, whereby all Maruyoshi-Song flows were seen to originate from certain T-brane deformations \cite{Heckman:2010qv,Cecotti:2010bp} of the Weierstrass geometry in F-theory. To this end, rather than aiming for a general treatment, we find it more convenient to work with a simple class of SCFT's. Extrapolating the rules of our approach to treat more complicated theories (in particular any linear quiver) can be done straightforwardly.

\begin{figure}[h]
\centering
\subfigure[]{\label{fig:Neq2}
\includegraphics[width=7cm]{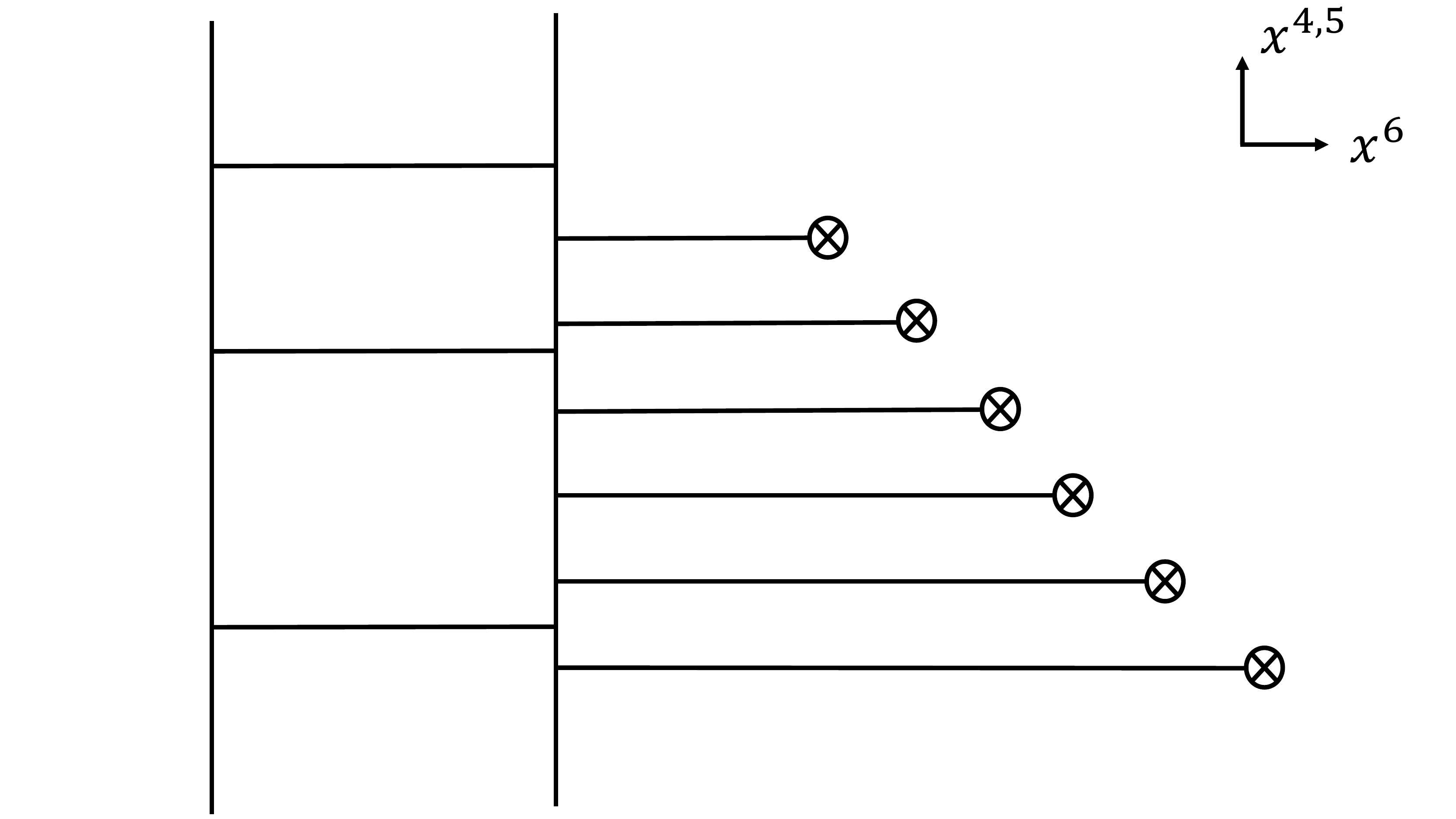}}
\subfigure[]{\label{fig:Neq1}
\includegraphics[width=7cm]{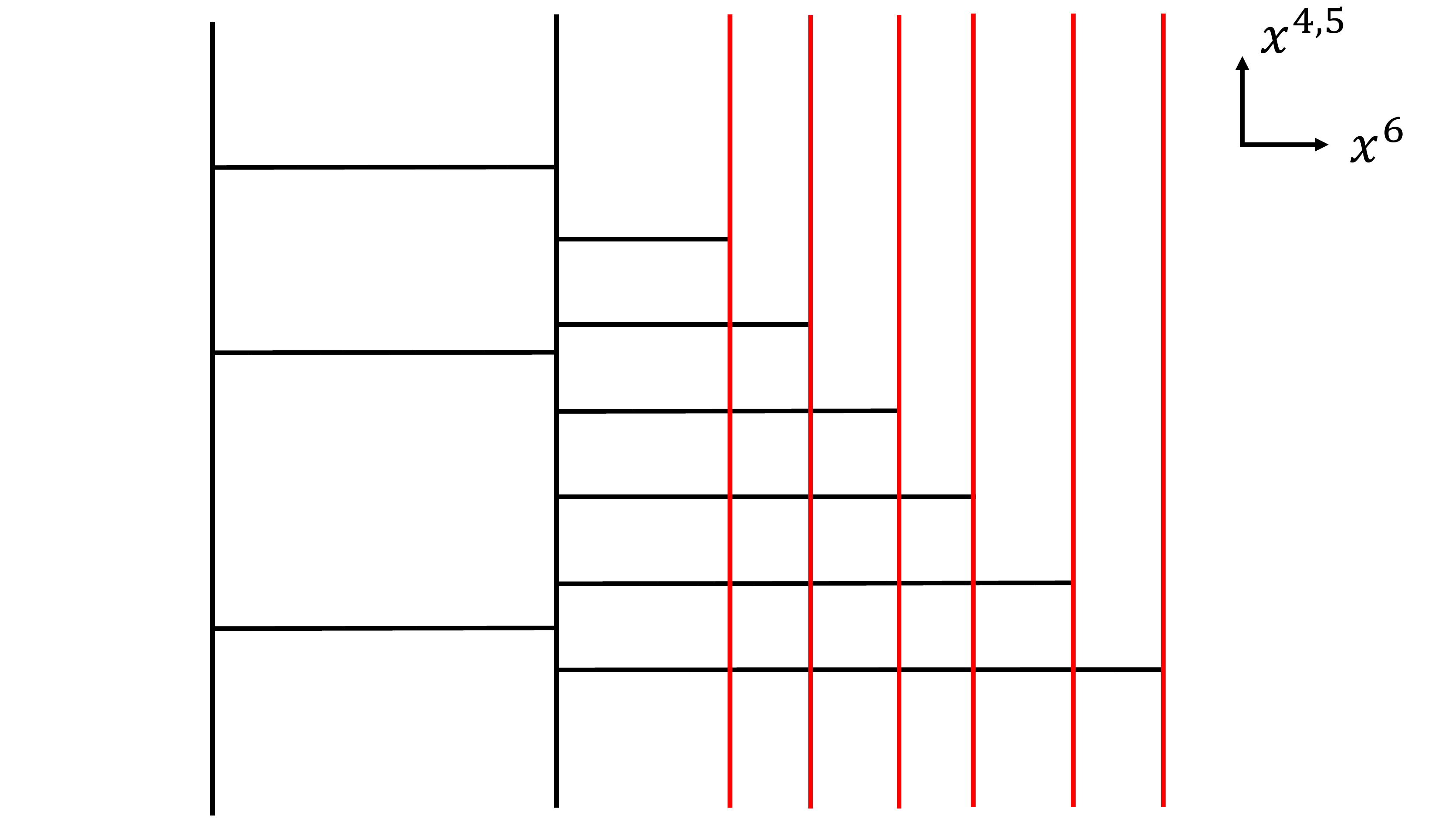}}
\caption{(a):  Brane configuration for 4d $\mathcal{N}=2$ SQCD. The figure shows the case with $N=3$. $\otimes$ represents an $\cN=2$ D6-brane. (b).  Brane configuration obtained after rotating the $\cN=2$ D6-brane in Figure \ref{fig:Neq2}. The $\cN=1$ D6-branes are depicted as red vertical lines.}
\label{fig:branesqcd}
\end{figure}

Consider 4d $\cN=2$ SQCD with $N$ colors and $2N$ flavors as the starting theory. Its SW geometry can be easily derived from a standard Witten cartoon involving D4, D6 and NS5-branes \cite{Witten:1997sc} (see Figure \ref{fig:Neq2}). The various branes extend in ten-dimensional flat space as follows:

\begin{center}
$
\begin{array}{c|cccccccccc}
\text{Witten cartoon}& 0 & 1 & 2 & 3 & 4 & 5 & 6 & 7 & 8 & 9 \\  \hline 
 \text{D4} & \times & \times & \times &  \times &  &  & \times &  &    \\
\text{NS5}  & \times & \times & \times & \times & \times  & \times &  &  &  &   \\
\text{D6$_{\cN=2}$}  & \times & \times & \times & \times &   &  &  & \times & \times &  \times \\ \hline 
\text{D6$_{\cN=1}$}  & \times & \times & \times & \times &\times   &  \times&  & \times &  &   
\end{array}
$
\end{center}
where the subscripts $_{\cN=2}$ and $_{\cN=1}$ indicate the amount of supersymmetry preserved by the orientation of the corresponding D6-branes. To engineer $\cN=2$ SQCD, $\cN = 2$ D6-branes can be placed in any place and different configuraitons are related by the Hanany-Witten transitions \cite{Hanany:1996ie}. Here we have chosen to work in the Hanany-Witten frame where all of the $\cN=2$ D6-branes are on one side of the two NS5-branes, which makes the whole $U(2N)$ flavor symmetry manifest, and thus allows us to access all of its nilpotent orbits when turning on the deformation. The SW geometry shows up in the M-theory uplift as the internal world-volume of the M5-brane lifting the above D4/NS5 configuration. This spans a holomorphic curve within the Taub-NUT space lifting the D6$_{\cN=2}$, and it has the following general form
\bea\label{GeneralXn}
z^2+c_1p_N(x)z+c_2\det\left(x\mathbb{1}_{2N}-M\right)=0\,,\qquad \lambda=x\frac{\d z}{z}\,,
\eea
where $z$ is a coordinate combining direction $6$ and M-theory circle, $x$ stands for directions $4,5$, $c_1,c_2$ are constants depending on the exactly marginal gauge coupling, $p_N(x)$ is a monic polynomial in $x$ of degree $N$, encoding the $N-1$ CB parameters in its coefficients, $M$ is the mass matrix for the $U(2N)$ flavors, and $\lambda$ indicates the SW differential. From this perspective, the eigenvalues of $M$ are understood as the relative position of the $N$ D6 branes in the directions $4$ and $5$ with respect to the stack of $N$ D4-branes connecting the two NS5-branes.

The $\cN=2\to\cN=1$ coupling \eqref{MSdef} can now be implemented simply by rotating all of the D6-branes and taking them oriented like the D6$_{\cN=1}$ in the table above \cite{Giveon:1998sr, Xie:2013gma}. Recalling that the meson $\mu$ is made of fields originating from the strings stretching between the $N$ gauge D4-branes and the $2N$ flavor D4-branes, this rotation has the effect of promoting the mass matrix $M$ to a 4d dynamical chiral field, describing the now free motion of the flavor D4-branes in directions $4,5$, which are now shared by NS5 and D6-branes. The $\cN=1$ brane configuration is depicted in Figure \ref{fig:Neq1}. 
Therefore, in order to study all Maruyoshi-Song flows of SQCD, it will suffice to insert in \eqref{GeneralXn} the explicit form of the ``flipping'' field $M$
\be\label{Mdef}
M=\rho(\sigma^+)+\sum_j M_{j,-j}\,,
\ee
where $\rho$ indicates the nilpotent embedding and $ M_{j,-j}$ the fluctuation associated to the lowest component of the spin $j$ representation of the embedded $SU(2)$. The sum extends over all spins appearing in the decomposition of the adjoint representation of $SU(2N)$ (see \cite{Gadde:2013fma, Agarwal:2014rua}).

The logic just described is completely general and can be applied to any starting SCFT in 4d, even non-Lagrangian ones: The space \eqref{GeneralXn}, which we dubbed $X_n$ in the introduction, has the general structure of a genus-$r$ Riemann-surface fibered over a base of dimension $n-r>r$, where $r$ is the rank of the theory. Studying whether a given orbit leads to SUSY enhancement is reduced to analyzing whether near the origin the fibration structure of $X_n$ is non-trivial only on a $r$-dimensional base.

This picture nicely connects to the description of the enhancement via Hitchin systems we discussed in Section \ref{s:Hitchin}. The Witten cartoon we have seen for SQCD translates into a class-$\cS$ configuration characterized by a two-sphere with one regular maximal puncture, carrying $SU(2N)$ flavor symmetry, and one irregular puncture, accounting for the two ``unbalanced'' NS5-branes of Figure \ref{fig:Neq2}. The SW geometry, in turn, arises as a $2N$-branched cover of the punctured sphere \cite{Gaiotto:2009hg}
\be\label{P_Phi}
\det\left(\lambda\mathbb{1}_{2N}-\Phi(z)\right) = 0\,,
\ee
where $z$ is the local coordinate on the sphere, $x$ is the local fiber coordinate of its canonical bundle, and $\Phi$ is the Hitchin field, a meromorphic section of $\cO(-2)$ on the sphere, with poles at the punctures. By reducing this configuration back to type IIA along a different circle, say direction $3$, $\Phi$ acquires the interpretation of the field of transverse deformations along directions $4,5$ of a stack of $2N$ D4-branes wrapped on the punctured sphere:

\begin{center}
$
\begin{array}{c|cccccccccc}
\text{class-$\cS$}& 0 & 1 & 2  & 4 & 5 & 6 & 7 & 8 & 9 &10 \\  \hline 
 \text{D4} & \times & \times & \times &   &  &  \times &  &  & & \times   \\
\text{$\Phi$}  &  &  &  & \times & \times  &  &  &  &  &   \\
\text{$\Phi^\prime$}  &  &  &  &  &   &  &  & \times & \times &   
\end{array}
$
\end{center}

In this different duality frame, the r\^ole of the mass matrix $M$ is played by a second Hitchin field $\Phi^\prime$, constant over the sphere, representing the transverse deformation of the D4 stack along directions $8,9$. This is because $\Phi^\prime$ is identified with the complex scalar in the vector multiplet of the three-dimensional $\cN=4$ theory living on the stack, and as such it couples to the matter $\mu$ localized at the regular puncture as $\Tr(\mu\Phi^\prime)$ \cite{Yonekura:2013mya}. As explained in Section \ref{s:Hitchin}, activating the SUSY breaking deformation \eqref{MSdef}, therefore, translates in this context to promoting $\Phi^\prime$ to a meromorphic section\footnote{In Section \ref{s:Hitchin} this meromorphic section was called $\Phi_1$.} of $\cO(-1)$, and to viewing the $\cN=1$ geometry $X_n$ (Eq.~\eqref{GeneralXn} with $M$ regarded as a field) as the intersection of \eqref{P_Phi} with the second spectral equation
\be\label{P_Phi'}
\det\left(\lambda^\prime\mathbb{1}_{2N}-\Phi^\prime(z)\right) = 0\,.
\ee
Recall that $\Phi^\prime$ is taken completely smooth at the regular puncture $z=0$. At this location, as is evident from Eq. \eqref{GeneralXn}, $\Phi^\prime$ has exactly the same spectral data of $M$, thus elucidating the meaning of the flipping field within the generalized Hitchin system of $\cN=1$ class-$\cS$ theories.

\subsection{SQCD with $6$ flavors}\label{subs:Su3}

With these geometric discussions in mind, we now analyze in detail a 4d $\mathcal{N}=2$ SQCD with $N=3$, and the systematics of its Maruyoshi-Song flows. We first focus on cases where the supersymmetry is enhanced to $\mathcal{N}=2$ at IR. In order to see if the resulting curve describes a 4d $\mathcal{N}=2$ superconformal field theory, we will make use of two necessary conditions which are satisfied for an $\mathcal{N}=2$ superconformal field theory. The first condition is that the genus of the curve should agree with the number of CB operators. The second condition is that if there is a parameter $a$ with $1 < D(a) \leq 2$ in the curve, then there should be only one parameter $b$ which satisfies $D(a) + D(b) = 2$. Since these are necessary conditions we cannot say exactly that the curve satisifying the two conditions describes an $\mathcal{N}=2$ superconformal field theory. We can only say that it is not inconsistent that it does. However the two conditions are more powerful when we single out theories that do not lead to supersymmetry enhancement. Indeed, in Subsection \ref{ExNotEnhance}, we will see cases which do not satisfy at least one of the two conditions, and hence the supersymmetry is not enhanced for those cases.

We first start from the SW curve of 4d $\mathcal{N}=2$ $SU(3)$ gauge theory with six flavors. The explicit form of the curve can be obtained from \eqref{GeneralXn} with $N=3$ and it is given by 
\begin{align}
z^2 + \left(a_1x^3 + a_2x+ a_3\right)z + \prod_{i=1}^{6}\left(x-m_i\right)=0\,, \label{SWSU3}
\end{align}
where we chose 
\begin{align}
M = \text{diag}(m_1, m_2, m_3, m_4, m_5, m_6)\,.\label{massSU3}
\end{align}
Here $\text{diag}(a, b, c, \cdots)$ denotes a diagonal maxtrix with the entries $a, b, c, \cdots$ and $m_i, (i=1, \cdots, 6)$ are mass parameters for the six flavors. The SW diffrential is the same as the general form in \eqref{GeneralXn}, namely, 
\begin{align}
\lambda=\frac{x}{z}dz\,. \label{lambdaSW}
\end{align}
Note that, by a $z$-dependent shift of $x$, the SW differential changes only by a total derivative term. We used a constant shift of $x$ to eliminate the monomial $x^2z$ in \eqref{SWSU3}. We have also rescaled $x$ to get rid of the overall constant in front of the product term.

We can determine the scaling dimension of the parameters $a_1, a_2, a_3$ in \eqref{SWSU3} from the fact that the scaling dimension of the SW differential \eqref{lambdaSW} is equal to one. This fixes the dimension of $x$ to be $1$. Then, from homogeneity of the curve polynomial \eqref{SWSU3}, the dimension of $z$ is $3$ and we have
\begin{align}
D^{\rm UV}(a_1) = 0\,, \qquad D^{\rm UV}(a_2) = 2\,, \qquad D^{\rm UV}(a_3) = 3\,,
\end{align}
where the superscript $^{\rm UV}$ reminds us that this is the UV theory. We can then interpret $a_1$ as the coupling constant and $a_2, a_3$ as CB operators.

We can see that the curve \eqref{SWSU3} satisfies the two necessary conditions for a 4d $\mathcal{N}=2$ superconformal field theory. The equation \eqref{SWSU3} describes a genus-two curve at a generic point on the Coulomb-branch moduli space. Since we have two CB operators $a_2$ and $a_3$, the genus of the curve indeed agrees with the number of the CB operators. Regarding the second condition, we have one parameter $a_2$ which satisfies $1 < D^{\rm UV}(a_2) \leq 2$. Then we can see that the curve contains the associated parameter $a_1$ which satisfies $D^{\rm UV}(a_1) + D^{\rm UV}(a_2) = 2$.

As described in Subsection \ref{subs:Prel}, when we turn on the $\mathcal{N}=1$ coupling with an adjoint chiral multiplet, the equation of the curve is essentially the same as \eqref{SWSU3} but the mass matrix $M$ is now promoted to a dynamical chiral field. Namely, we consider the curve
\begin{align}
z^2 + \left(a_1x^3  + a_2x+ a_3\right)z + \det\left(x\mathbb{1}_6 - M -\tilde{M}_1\mathbb{1}_6\right)=0\,, \label{Neq1SW1}
\end{align} 
where $M$ is given by \eqref{Mdef} and $\tilde{M}_1\mathbb{1}_6$ corresponds to the trace component. Note that the SW differential is not necessarily the same as \eqref{lambdaSW}, and needs to be determined for each example.

In this subsection we consider two examples, i.e.~the nilpotent orbits $[6]$ and $[5, 1]$. It is known from a-maximization that these cases lead to supersymmetry enhancement  \cite{Maruyoshi:2016aim, Agarwal:2016pjo}. Here, instead, we carry out this analysis in a purely algebraic manner, using the $\cN=1$ curve \eqref{Neq1SW1}.

\subsubsection*{Orbit $[6]$ of $SU(6)$}\label{maximalsection}
We first consider turning on a vev in the maximal nilpotent orbit of $SU(6)$, labeled by $[6]$. For this nilpotent orbit, the raising operator of the $\mathfrak{sl}(2)$ standard triple\footnote{For a standard reference on building standard triples see \cite{CMcG}.} is canonically defined to be:
\begin{align}
\rho(\sigma^+) = \left(
\begin{array}{cccccc}
0 & 1 & 0 & 0 & 0 & 0\\
0& 0 & 1 & 0 & 0 & 0\\
0 & 0 & 0 & 1 & 0 & 0\\
0 & 0 & 0 & 0 & 1 & 0\\
0 & 0 & 0 & 0 & 0 & 1\\
0 & 0 & 0 & 0 & 0 & 0
\end{array}
\right).\label{maximal}
\end{align}
Under the background \eqref{maximal} the adjoint representation of the $\mathfrak{su}(6)$ flavor algebra splits according to the branching rule 
\begin{equation}
\mbox{adj}\to V_{5}\oplus V_4\oplus V_3\oplus V_2\oplus V_1 \label{branchmax}\,,
\end{equation}
where with $V_j$ we denote the $\mathfrak{sl}(2)$ irreducible representation of spin $j$.
The components that remain coupled are the lowest components of each spin $j$ representation, namely $M_{j, -j}$ for $j=1, \cdots, 5$. Hence the $M$ in \eqref{Neq1SW1} is given by
\begin{align}
M = \left(
\begin{array}{cccccc}
0 & 1 & 0 & 0 & 0 & 0\\
5 M_{1, -1} & 0 & 1 & 0 & 0 & 0\\
5 M_{2, -2} & 8 M_{1, -1} & 0 & 1 & 0 & 0\\
5 M_{3, -3} & 9 M_{2, -2} & 9 M_{1, -1} & 0 & 1 & 0\\
 M_{4, -4} & 8 M_{3, -3} & 9 M_{2, -2} & 8 M_{1, -1} & 0 & 1\\
M_{5, -5} &  M_{4, -4} & 5 M_{3, -3} & 5 M_{2, -2} & 5 M_{1, -1} & 0
\end{array}
\right).\label{Mdefmaximal}
\end{align}

Then the characteristic polynomial in the curve \eqref{Neq1SW1} becomes 
\begin{equation}\label{detmaximal}
\begin{aligned}
\det(x\mathbb{1}_6 - M -\tilde{M}_1\mathbb{1}_6)=&x^6 -6\tilde{M}_1x^5+\left(15\tilde{M}_1^2- 35 M_{1,-1}\right)x^4+\\
&+\left(-20\tilde{M}_1^3+140\tilde{M}_1M_{1,-1} - 28 M_{2,-2}\right)x^3 +\\
&+ \left(15\tilde{M}_1^4-210\tilde{M}_1^2M_{1,-1}+259 M_{1,-1}^2+84\tilde{M}_1M_{2,-2} - 18 M_{3,-3}\right)x^2 +\\
&+ \left(-6\tilde{M}_1^5+140\tilde{M}_1^3M_{1,-1}+-518\tilde{M}_1M_{1,-1}^2-84\tilde{M}_1^2M_{2,-2}+\right.\\
&\left.+220 M_{1,-1}  M_{2,-2}+36\tilde{M}_1M_{3,-3} - 2  M_{4,-4}\right)x+\\
&+\tilde{M}_1^6-35\tilde{M}_1^4M_{1,-1}+259\tilde{M}_1^2M_{1,-1}^2-225 M_{1,-1}^3+28 \tilde{M}_1^3M_{2,-2}+\\
&+220\tilde{M}_1M_{1,-1}M_{2,-2}+ 25 M_{2,-2}^2-18\tilde{M}_1^2M_{3,-3} + 50 M_{1,-1} M_{3,-3}+\\
&+2\tilde{M}_1M_{4,-4}- M_{5,-5} \,.
\end{aligned}
\end{equation}
We can redefine $ M_{j, -j}, (j=1, \cdots, 5)$ and $\tilde{M}_1$ to rewrite \eqref{detmaximal} as
\begin{align}
\det(x\mathbb{1}_6 - M -\tilde{M}_1\mathbb{1}_6)=x^6 +M_1x^5+ M_2x^4 + M_3x^3 + M_4x^2 + M_5x +M_6\,,
\end{align}
where the $M_i$, $i=2,\cdots 6$ are now the Casimir invariants of $SU(6)$. In the end we arrive at the equation 
\begin{align}
z^2 + \left(a_1x^3 + a_2x+ a_3\right)z + x^6 +M_1x^5+ M_2x^4 + M_3x^3 + M_4x^2 + M_5x +M_6 = 0\,, \label{SWmaximal}
\end{align}
after turning on the $\mathcal{N}=1$ deformation \eqref{Mdefmaximal}.

We now interpret \eqref{SWmaximal} as the IR curve after the RG flow. Since the curve \eqref{SWmaximal} is essentially the same equation as \eqref{SWSU3}, it will have the same holomorphic one-forms. On the other hand, the holomorphic one-forms can be obtained by taking a derivative of the SW differential with respect to CB operators. In order to determine the scaling dimension of the parameters in \eqref{SWmaximal}, we use that relative scaling dimensions are RG-flow invariant, and thus that the holomorphic one-form corresponding to the UV CB operator with the maximal dimension, i.e.~$a_3$, is the same as the holomorphic one-form associated with the IR CB operator with the maximal scaling dimension, which we postulate to be $M_6$. In other words, the new SW differential in the IR, $\lambda_{[6]}$, must satisfy
\begin{align}
\frac{\d\lambda}{\d a_3} = \frac{\d \lambda_{[6]}}{\d M_6}\,, \label{ansatzmaximal}
\end{align}
where $\lambda$ is given by \eqref{lambdaSW}. Condition \eqref{ansatzmaximal} leads to the following relation between parameters and coordinates 
\begin{align}
D^{\rm IR}(z) - 5D^{\rm IR}(x) = 1 - D^{\rm IR}(M_6)\,.
\end{align}
Equation \eqref{SWmaximal} also implies $D(z) = 3D(x)$ and $2D(z) = D(M_6)$\footnote{The symbol $D$ without subscript indicates the scaling dimension anywhere along the RG flow.}. Then, the scaling dimension of the parameter $M_6$ can be fixed as 
\begin{align}
D^{\rm IR}(M_6) = \frac{3}{2}\,.
\end{align}
The scaling dimension of the other parameters can also be determined: 
\begin{equation}
\begin{split}
&D^{\rm IR}(M_5) = \frac{5}{4}\,, \quad D^{\rm IR}(M_4) = 1\,, \quad D^{\rm IR}(M_3) = \frac{3}{4}\,, \quad D^{\rm IR}(M_2) = \frac{1}{2}\,, \quad D^{\rm IR}(M_1) = \frac{1}{4}\,, \quad\\
& D^{\rm IR}(a_1) = 0\,, \quad D^{\rm IR}(a_2) = \frac{1}{4}\,, \quad D^{\rm IR}(a_3) = \frac{1}{2}\,, \quad D^{\rm IR}(a_4) = \frac{3}{4}\,.
\end{split}
\end{equation}
Therefore the parameters $M_6$ and $M_5$ may be identified as the two CB operators in the IR, being the only ones whose dimension is strictly above the unitarity bound.

Let us see if the resulting curve \eqref{SWmaximal} satisfies the two necessary conditions for an $\mathcal{N}=2$ superconformal field theory. First, the genus of the curve \eqref{SWmaximal} is two and the number of the CB operators is also two. Therefore the first condition is satisfied. For the second condition, we need to see carefully if we can eliminate any parameters in \eqref{SWmaximal} by a change of coordinates which leaves the SW differential invariant up to a total derivative. For that we need to determine the SW differential $\lambda_{[6]}$ explicitly for the IR theory.

Note that we can write the SW differential as $\lambda_{[6]} = f(z, x(z,a))dz$,\footnote{Here we are assuming that the SW differential does not have an explicit dependence on $a$. In the next subsection, we will analyze an example where we will need to relax this hypothesis.} where $a$ is any CB parameter and $x$ is regarded as a function of $z$ from the curve equation \eqref{SWmaximal}, $F(z, x, a) = 0$. Hence, the derivative of $\lambda_{[6]}$  with respect to $a$ can be written as 
\begin{align}\label{SWderivative}
\frac{d \lambda}{d a} = -\frac{\partial f(z, x)}{\partial x}\frac{\partial F(z, x, a)}{\partial a}\left(\frac{\partial F(z, x, a)}{\partial x}\right)^{-1}dz\,,
\end{align}
Using \eqref{SWderivative}, the relation \eqref{ansatzmaximal} implies that
\begin{align}
\frac{d \lambda_{[6]}}{d x} = dz\,,
\end{align}
which we can trivially solve, leading to
\begin{align}
\lambda_{[6]} = xdz\,, \label{diffmaximal}
\end{align}
up to total-derivative terms.

As opposed to the SW differential in the UV, Eq. \eqref{lambdaSW}, the one in the IR, Eq. \eqref{diffmaximal}, crucially allows for a new change of coordinates: an $x$-dependent shift of $z$. It is also possible to shift $x$ by a constant. Therefore, Eq. \eqref{SWmaximal} can be further simplified by eliminating the term linear in $z$ and the term proportional to $x^5$: 
\begin{align}
z^2 + x^6 + M_2x^4 + M_3x^3 + M_4x^2 + M_5x +M_6 = 0\,. \label{SWmaximal2}
\end{align}
It is now possible to see that the second condition is indeed satisfied. Namely we have two pairs of coupling constant/CB operator, $(M_2,M_6)$ and $(M_3,M_5)$, which satisfy 
\begin{align}
D^{\rm IR}(M_2) + D^{\rm IR}(M_6) = 2\,, \qquad D^{\rm IR}(M_3) + D^{\rm IR}(M_5) = 2\,.
\end{align}
Finally, $M_4$ plays the r\^ole of mass parameter for the IR flavor symmetry.

Since the curve \eqref{SWmaximal2} with the SW differential \eqref{diffmaximal} satisfies the two conditions, the theory described by the curve is compatible with an $\mathcal{N}=2$ superconformal field theory. Indeed in this case we know that the curve is nothing but the SW curve of the $(A_1, A_5)$ generalized Argyres-Douglas theory, which has $U(1)$ flavor symmetry. This is consistent with the result of \cite{Maruyoshi:2016aim}.

\subsubsection*{Orbit $[5,1]$ of $SU(6)$}

Let us now consider turning on a vev for the subregular nilpotent orbit of $SU(6)$. Such an orbit is labeled by the partition $[5,1]$ of the number $N_f=6$. For this nilpotent orbit, the raising operator of the $\mathfrak{sl}(2)$ standard triple is canonically defined to be

\begin{equation}
\rho(\sigma^+)=\left(\begin{array}{cccccc}
$0$ & $1$ & $0$ & $0$ & $0$ & $0$ \\ 
$0$ & $0$ & $1$ & $0$ & $0$ & $0$ \\ 
$0$ & $0$  & $0$ & $1$ & $0$ & $0$ \\ 
$0$ & $0$ & $0$ & $0$ & $1$ & $0$ \\ 
$0$ & $0$ & $0$ & $0$ & $0$ & $0$ \\ 
$0$ & $0$ & $0$ & $0$ & $0$ & $0$
\end{array} \right).
\label{triple51}
\end{equation}

Under the background \eqref{triple51} the adjoint representation of the $\mathfrak{su}(6)$ flavor algebra splits according to the branching rule 
\begin{equation}
\mbox{adj}\to V_{4}\oplus V_3\oplus 3V_2\oplus V_1\oplus V_0 \label{branchsubreg}\,,
\end{equation}
where with $V_j$ we denote the $\mathfrak{sl}(2)$ irreducible representation of spin $j$.
As usual, the components of the field $M$ that remain coupled after turning on the vev (\ref{triple51}) are given by the lowest component of each $\mathfrak{sl}(2)$ spin $j$ representation appearing in (\ref{branchsubreg}), namely $M_{j, -j}$ for $j=1, \cdots, 5$. Hence the $M$ in \eqref{Neq1SW1} is given in this case by
\begin{align}
M = \left(
\begin{array}{cccccc}
M_{0,0} & 1 & 0 & 0 & 0 & 0\\
2M_{1,-1} & M_{0,0} & 1 & 0 & 0 & 0\\
2\sum_{i=1}^3M_{2,-2}^{(i)} & 3M_{1,-1} & M_{0,0} & 1 & 0 & 0\\
M_{3,-3} & 3\sum_{i=1}^3M_{2,-2}^{(i)}  & 3M_{1,-1} & M_{0,0} & 1 & 0\\
M_{4,-4} & M_{3,-3} & 2\sum_{i=1}^3M_{2,-2}^{(i)} & 2M_{1,-1} & M_{0,0} & 2M^{(2)}_{2,-2}\\
2M^{(3)}_{2,-2} &  0 & 0 & 0 & 0 & -5M_{0,0}
\end{array}
\right).\label{subregularorb}
\end{align}
where we have denoted with $M^{(i)}_{2,-2}$, $i=1,2,3$ the lowest spin component of the three different $V_2$ representations appearing in \eqref{branchsubreg}.

Now we need to compute the characteristic polynomial of the matrix $M-\mathbb{1}_6\tilde{M}_1$, and perform an analysis analog to that of section (\ref{maximalsection}). However, for ease of presentation of the result, let us just write the characteristic polynomial of $M$, and re-install the trace part at a later stage. Such characteristic polynomial can be computed as 
\begin{equation}
\begin{aligned}
\det\left(\mathbb{1}_6x-M\right)&=x^6-x^4 \left(15 M_{0,0}^2+10 M_{1,-1}\right)+x^3 \left(40 M_{0,0}^3-20 M_{0,0}
M_{1,-1}-7 \sum_{i=1}^3M_{2,-2}^{(i)}\right)+\\&+x^2
\left(-45 M_{0,0}^4+120 M_{0,0}^2 M_{1,-1}-21 M_{0,0} \sum_{i=1}^3M_{2,-2}^{(i)}+16
M_{1,-1}^2-2 M_{3,-3}\right)+\\
&x \left(24 M_{0,0}^5-140 M_{0,0}^3 M_{1,-1}+63
M_{0,0}^2 \sum_{i=1}^3M_{2,-2}^{(i)}+\right.\\
&+\left.64 M_{0,0} M_{1,-1}^2-8 M_{0,0} M_{3,-3}+8 M_{1,-1}
\sum_{i=1}^3M_{2,-2}^{(i)}-M_{4,-4}\right)+\\
&+40 M_{0,0} M_{1,-1} \sum_{i=1}^3M_{2,-2}^{(i)}-5 M_{0,0}
M_{4,-4}-4 M_{(2,-2)}^{(2)} M_{(2,-2)}^{(3)}+\\
&-5 M_{0,0}^6+50 M_{0,0}^4 M_{1,-1}-35 M_{0,0}^3 \sum_{i=1}^3M_{2,-2}^{(i)}-80 M_{0,0}^2
M_{1,-1}^2+10 M_{0,0}^2 M_{3,-3}\,.
\end{aligned}
\end{equation}
We can now redefine the singlets as follows
\begin{equation}
\begin{aligned}
M_2&:=15 M_{0,0}^2+10 M_{1,-1}\,,\\
M_3&:=40 M_{0,0}^3-20 M_{0,0}
M_{1,-1}-7 \sum_{i=1}^3M_{2,-2}^{(i)}\,,\\
M_4&:=-45 M_{0,0}^4+120 M_{0,0}^2 M_{1,-1}-21 M_{0,0} \sum_{i=1}^3M_{2,-2}^{(i)}+16
M_{1,-1}^2-2 M_{3,-3}\,,\\
M_5&:=24 M_{0,0}^5-140 M_{0,0}^3 M_{1,-1}+63
M_{0,0}^2 \sum_{i=1}^3M_{2,-2}^{(i)}+\\&+64 M_{0,0} M_{1,-1}^2-8 M_{0,0} M_{3,-3}+8 M_{1,-1}
\sum_{i=1}^3M_{2,-2}^{(i)}-M_{4,-4}\,.
\end{aligned}
\end{equation}

The characteristic polynomial in terms of these new variables can be written in a much more compact form, namely
\begin{equation}
\det\left(\mathbb{1}_6x-M-\mathbb{1}_6\tilde{M}_1\right)=x^6+M_1x^5+M_2x^4+M_3x^3+M_4x^2+M_5x+M_6\,,
\label{chpoly51}
\end{equation}
where we have re-installed the trace part, and we have defined
\begin{equation}\label{M6}
\left.M_6\right|_{\tilde{M}_1=0}:=-15625 M_{0,0}^6+625 M_{0,0}^4 M_2+125 M_{0,0}^3 M_3-25
M_{0,0}^2 M_4+5 M_{0,0} M_5-4 M_{2,2}^{(2)} M_{2,2}^{(3)}\,.
\end{equation}
We stress that while in equation (\ref{chpoly51}) the quantities $M_2$, $\cdots$, $M_5$ have to be considered independent variables, $M_6$ is instead explicitly dependent on all of the $M_i$'s and also on $M_{0,0}$, $M_{2,-2}^{(2)}$, $M_{2,-2}^{(3)}$ and $\tilde{M}_1$.

The $\mathcal{N}=1$ deformed curve is therefore given by 
\begin{equation}
\begin{aligned}
&z^2 + \left(a_1x^3 + a_2x+ a_3\right)z + x^6 +M_1x^5+ M_2x^4 + M_3x^3 + M_4x^2 + M_5x+M_6
= 0\,, \label{SWSubreg}
\end{aligned}
\end{equation}

As in Subsection (\ref{maximalsection}), our strategy to compute the dimension of all the operators and couplings entering the SW curve of the infrared theory consists in making an Ansatz for the parameter playing the r\^ole of the IR CB operator with the highest dimension. Since relative dimensions are RG-flow invariant, we are led to identify the highest spin $M_5$ as such operator. Then equating the holomorphic one-forms associated to the UV and IR CB operators of highest dimension, we get 
\begin{equation}
\dfrac{\d \lambda}{\d a_3}=\dfrac{\d \lambda_{[5,1]}}{\d M_5}\,,
\end{equation}
where we denoted by $\lambda_{[5,1]}$ the IR SW differential.
At the level of dimensions, this equation implies
\begin{equation}
D^{\rm IR}(z) - 5D^{\rm IR}(x) = 1 - D^{\rm IR}(M_5)\,. \label{dimop3}
\end{equation}
Now crucially equation (\ref{dimop3}), together with the homogeneity of the curve (\ref{SWSubreg}) fixes the dimensions of the coordinates $x$ and $z$ and the operator $M_5$ as
\begin{equation}
D^{\rm IR}(M_5) = \frac{5}{3}\,, \quad D^{\rm IR}(x)=\dfrac{1}{3}\,, \quad D^{\rm IR}(z)=1\,.
\end{equation}
This in turn fixes the dimensions of all the other parameters as
\begin{equation}
\begin{aligned}
D^{\rm IR}(M_4)&=\dfrac{4}{3}\,, \quad D^{\rm IR}(M_3)=1\,, \quad D^{\rm IR}(M_2)=\dfrac{2}{3}\,, \quad D^{\rm IR}(M_{0,0})=\dfrac{1}{3}\,, \\
D^{\rm IR}(a_1)&=0\,, \quad D^{\rm IR}(a_2)=\dfrac{1}{3}\,, \quad D^{\rm IR}(a_3)=\dfrac{2}{3}\,, \quad D^{\rm IR}(a_4)=1\,.
\end{aligned}
\end{equation}                                      

The parameters $M_{2,-2}^{(2)}$ and $M_{2,-2}^{(3)}$ only appear in the deformed curve (\ref{SWSubreg}) through their product, so it seems that we could only infer
\begin{equation}
D^{\rm IR}(M_{2,-2}^{(2)})+D^{\rm IR}(M_{2,-2}^{(3)})=2\,.
\end{equation}
However, we also know they must have the same dimension as they have the same spin under the Jacobson-Morozov $\mathfrak{sl}(2)$, as shown in (\ref{branchsubreg}). This is enough to conclude that
\begin{equation}
D^{\rm IR}(M_{2,-2}^{(2)})=D^{\rm IR}(M_{2,-2}^{(3)})=1\,.
\end{equation}
Nevertheless, these two as well as $M_{0,0}$ are not to be considered as independent parameters, since they only enter the low-energy effective theory through the combination $M_6$ \eqref{M6}.

We can therefore see that the parameters $M_5$ and $M_4$ can be identified as CB operators in the IR, as they are the only ones of dimension stricly greater than one.

Let us now check if the chosen Ansatz leads to a SW geometry that satisfies the two necessary conditions for an $\mathcal{N}=2$ superconformal field theory. We see that the genus of the curve (\ref{SWSubreg}) is two, and also the number of CB operators is two, so the first condition is satisfied. In order to check for the second condition, we need to eliminate any reduntant parameter in (\ref{SWSubreg}) by a coordinate trasformation which leaves the SW differential fixed (up to a total derivative).

In order to do this, we need first of all to solve for the SW differential of the IR theory. By using equation (\ref{SWderivative}) we find

\begin{equation}
\lambda_{[5,1]} = \frac{z}{x}\ dx = z\ d\log(x)\,, \label{SWdiff10}
\end{equation}
up to total-derivative terms. This form of the SW differential allows us to freely shift $z$ by a generic polynomial in $x$. We will use such a shift in order to reabsorb all the terms of the form $a_ix^{4-i}z$ for $i=1,\cdots 4$ in the curve (\ref{SWSubreg}).

In particular, the equation (\ref{SWSubreg}) can be thus simplified to 
\begin{equation}
\begin{aligned}
&z^2 + x^6 +M_1x^5+ M_2x^4 + M_3x^3 + M_4x^2 + M_5x+M_6= 0\,, \label{SWSubregsimp}
\end{aligned}
\end{equation}
where $D^{\rm IR}(M_6)=2$ and $D^{\rm IR}(M_1)=\frac{1}{3}$.

It is now possible to check that the second condition for the enhancement is indeed satisfied. Namely we have two pairs of coupling constant/CB operator, $(M_1,M_5)$ and $(M_2,M_4)$, satisfying
\begin{equation}
D^{\rm IR}(M_5)+D^{\rm IR}(M_1)=2\,, \qquad D^{\rm IR}(M_4)+D^{\rm IR}(M_2)=2\,.
\end{equation}
Finally $M_3$ and $M_6$ play the r\^ole of mass terms for the IR curve.

In conclusion we claim that, considering the orbit $[5,1]$, our method leads to a geometry specified by the curve (\ref{SWSubregsimp}) and the differential (\ref{SWdiff10}). Such a pair passes both our criteria of SUSY enhancement, so we expect that the $\mathcal{N}=1$ geometry has enhanced to $\mathcal{N}=2$. Our expectation is confirmed by the $a$-maximization analysis of \cite{Agarwal:2016pjo}. This flow is believed to land on the $(A_1, D_6)$ generalized Argyres-Douglas theory, which has $SU(2)\times U(1)$ flavor symmetry \cite{Xie:2012hs}. Indeed, as can be seen in (\ref{SWSubregsimp} and (\ref{SWdiff10}), our method naturally and explicitly reproduced the SW curve and differential of such a theory.

\subsection{Examples that do not enhance}\label{ExNotEnhance}

In Subsection \ref{subs:Su3}, we have considered the deformations corresponding to the orbits $[6]$ and $[5, 1]$ of $SU(6)$. Both cases satisfy the two conditions and they showed supersymmetry enhancement in the IR. In this section, we turn to cases that do not exhibit supersymmetry enhancement. One case is considering a different deformation in the 4d $SU(3)$ gauge theory with six flavors. In the other case we will use an SO-Sp quiver theory with the deformation corresponding to the maximal nilpotent orbit of the symplectic flavor symmetry, which was discussed in \cite{Giacomelli:2018ziv}. It is known that neither case leads to supersymmetry enhancement and we are going to confirm this claim using our algebraic criteria.

\subsubsection*{Orbit $[4,2]$ of $SU(6)$}
For the first case we consider a deformation with a vev corresponding to the nilpotent orbit labeled by $[4, 2]$. Namely the vev is given by 
\begin{align}
\rho(\sigma^+) = \left(
\begin{array}{cccccc}
0 & 1 & 0 & 0 & 0 & 0\\
0& 0 & 1 & 0 & 0 & 0\\
0 & 0 & 0 & 1 & 0 & 0\\
0 & 0 & 0 & 0 & 0 & 0\\
0 & 0 & 0 & 0 & 0 & 1\\
0 & 0 & 0 & 0 & 0 & 0
\end{array}
\right).\label{42}
\end{align}
The fluctuations around the background that remain coupled are as usual the lowest components of the spin $j$ representations and the matrix form of the deformation becomes
\begin{align}
M = \left(
\begin{array}{cccccc}
 M_0 & 1 & 0 & 0 & 0 & 0\\
-3 M_{1,-1} &  M_0 & 1 & 0 & 0 & 0\\
 M_{2,-2} & -4 M_{1,-1} &  M_0 & 1 &  M^{(1)}_{1, -1} & 0\\
- M_{3,-3} &  M_{2,-2} & -3 M_{1,-1} &  M_0 & - M^{(1)}_{2, -2} & 3 M^{(1)}_{1, -1}\\
 3 M^{(2)}_{1, -1} & 0 & 0 & 0 & -2 M_0 & 1\\
- M^{(2)}_{2, -2}  &  M^{(2)}_{1, -1}  & 0 & 0 & - M^{(3)}_{1, -1} & -2 M_0
\end{array}
\right).\label{Mdef42}
\end{align}
Inserting \eqref{Mdef42} into \eqref{Neq1SW1} yields the $\mathcal{N}=1$ curve.

Let us then determine the scaling dimension of the parameters appearing in the curve equation. Our Ansatz is that the CB operator with the highest dimension in the IR is $M_{3,-3}$. Denoting by $\lambda_{[4,2]}$ the IR SW differential, this yields the relation
\begin{align}
\frac{\d \lambda}{\d a_3} = \frac{\d \lambda_{[4,2]}}{\d  M_{3, -3}}\,,
\end{align}
which leads to
\begin{align}\label{relation42}
D^{\rm IR}(z) - 5D^{\rm IR}(x) = 1 - D^{\rm IR}( M_{3, -3})\,.
\end{align}
Then the relation \eqref{relation42} together with the curve equation fixes the scaling dimension of some of the parameters as 
\begin{align}
D^{\rm IR}(M_{3, -3}) = 2\,, \quad D^{\rm IR}(M_{2, -2}) = \frac{3}{2}\,, \quad D^{\rm IR}(M^{(3)}_{1, -1}) = 1\,, \quad D^{\rm IR}(M_0) = \frac{1}{2}\,.
\end{align}
Moreover, using that relative scalings are RG-flow invariant, we also find 
\begin{align}
D^{\rm IR}(M^{(1)}_{1, -1}) = D^{\rm IR}(M^{(2)}_{1, -1})= 1\,, \qquad D^{\rm IR}(M^{(1)}_{2, -2})= D^{\rm IR}(M^{(2)}_{2, -2}) &= \frac{3}{2}\,. \label{constraint42}
\end{align}
Hence $M_{3, -3}$ and $M_{2, -2}$ may serve as CB operators. But the relation \eqref{constraint42} implies that also $M^{(1)}_{2, -2}$ and $M^{(2)}_{2, -2}$ remain coupled in the IR. On the other hand, we know that the SW curve remains a genus-two curve throughout the flow. Therefore we now encounter a situation where the number of the CB operators does not agree with the genus of the curve, implying that the curve cannot describe an $\mathcal{N}=2$ superconformal field theory. This is consistent with the result of \cite{Maruyoshi:2016aim}.

\subsubsection*{An orthosymplectic quiver}
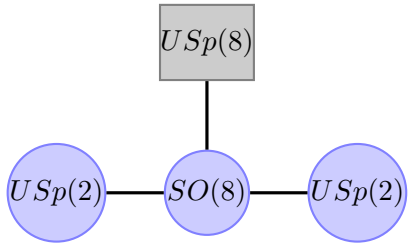
\begin{figure}[h]
	\begin{center}
		\begin{tikzpicture}
		\node (m1) at ( -2,0) [gauge] {$USp(2)$};
		\node (m2) at ( 0,0) [gauge] {$SO(8)$}
		edge [lpre] node[auto,swap] {} (m1);
		\node (m3) at (2,0) [gauge] {$USp(2)$}		
		edge [lpre] node[auto,swap] {} (m2);
		\node (m3) at (0,2) [global] {$USp(8)$}		
		edge [lpre] node[auto,swap] {} (m2);
		\end{tikzpicture}
		\caption{The orthosymplectic quiver.}
		\label{OrtoQuiver}
	\end{center}
\end{figure}

Here we want to apply a similar analysis to a Lagrangian theory with different gauge and flavor groups. The aim is to show an example that is known not to exhibit supersymmetry enhancement, but that nevertheless satisfies the criterion that the genus of the IR curve matches the dimension of the base over which it is fibered. Consider the quiver of Figure \ref{OrtoQuiver} where four flavors are attached to the $SO(8)$ gauge node. 

This theory was found in \cite{Giacomelli:2018ziv} to give no enhancement, because it violates an intricate relation imposed by 't Hooft anomaly matching, while preserving the rank. Using the underlying SW geometry, we would like to argue that the absence of enhancement originates from a mismatch between CB operators of dimension between $1$ and $2$ and coupling constants. This suggests an elegant geometric counterpart to the third criterion for enhancement discussed in Subsection 3.2 of \cite{Giacomelli:2018ziv}\footnote{The first two criteria in \cite{Giacomelli:2018ziv} are incorporated in our geometric condition that, if the theory is to preserve $\cN=2$ in the IR, the genus of the SW curve must be equal to the dimension of the base of the fibration.}.

\begin{figure}
\centering
\subfigure[]{\label{fig:brane1}
\includegraphics[width=7cm]{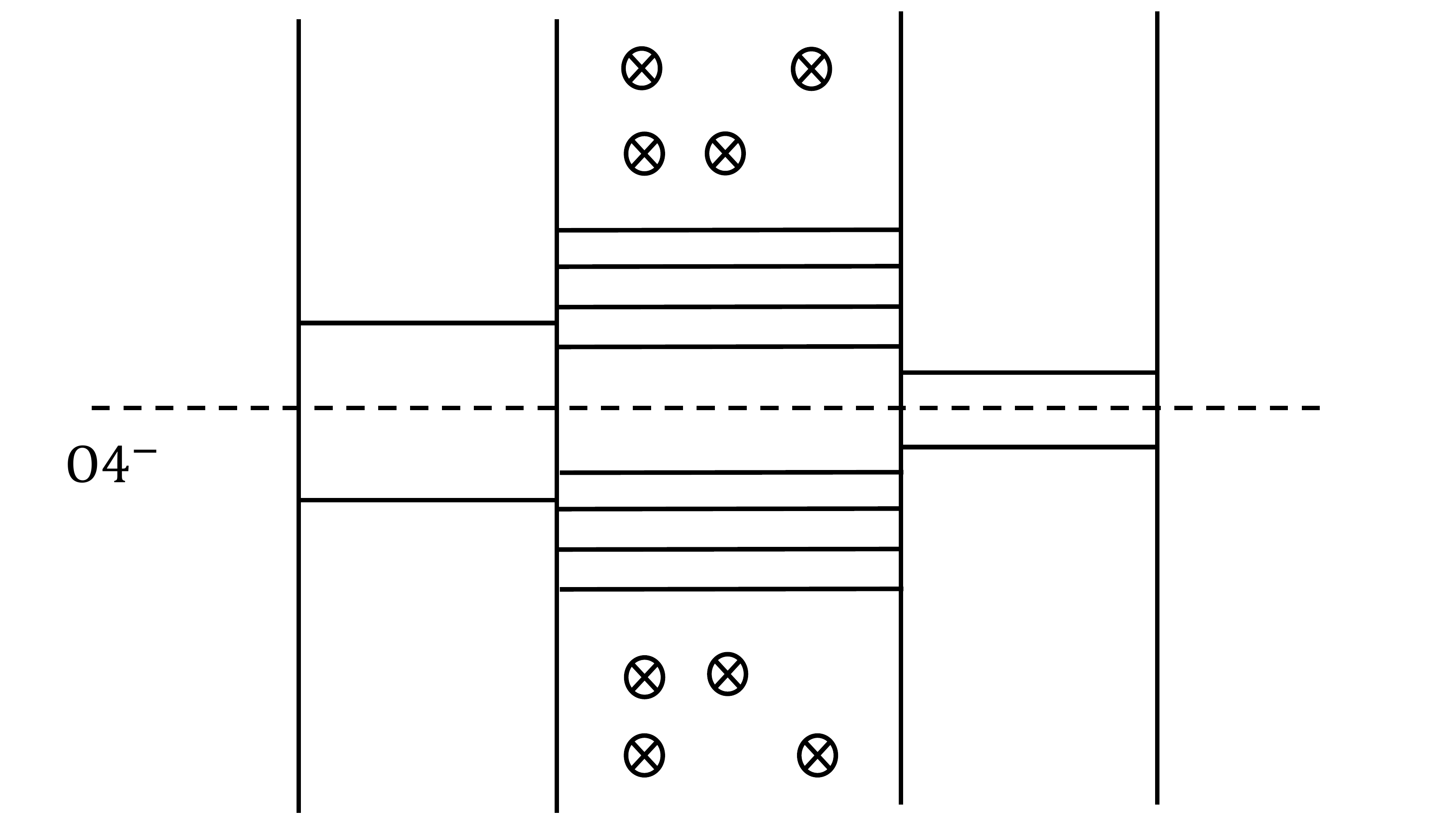}}
\subfigure[]{\label{fig:brane2}
\includegraphics[width=7cm]{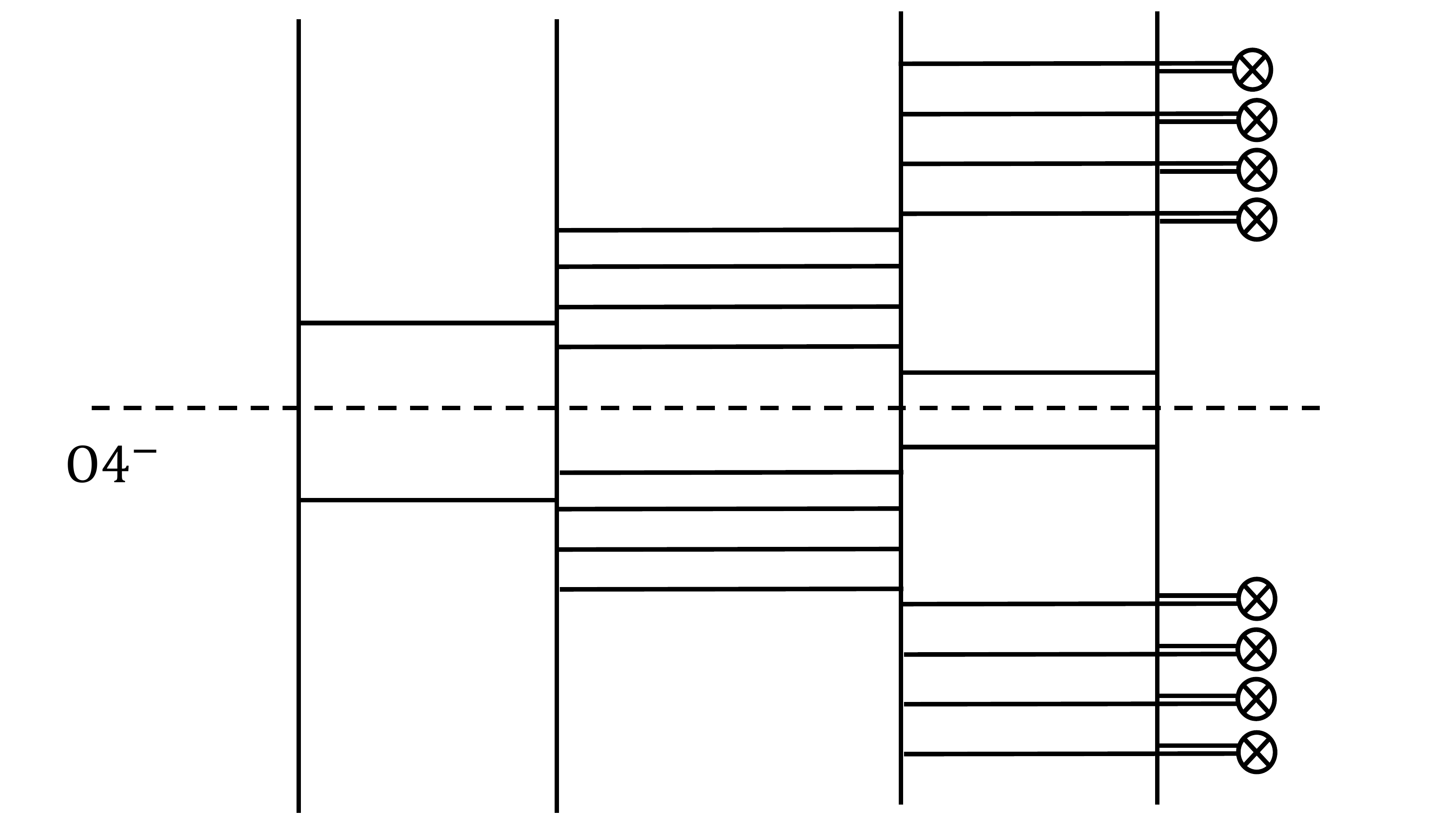}}
\caption{Brane picture realizing the orthosymplectic quiver of Figure \ref{OrtoQuiver}.}
\label{fig:brane}
\end{figure}

The SW curve of the theory can be obtained from a brane configuration realizing the quiver theory. This involves an O4-plane and the schematic picture is depicted in Figure \ref{fig:brane1}. 
The four D6-branes in the upper half-plane give four flavors to the $SO(8)$ gauge node. In order to read off the SW curve, it is useful to use a configuration which does not have any D6-branes. For that we move the D6-branes in between the middle NS5-branes in the right direction for example. A D4-brane is created when a D6-brane crosses an NS5-brane and the final configuration is given in Figure \ref{fig:brane2}. 

To write the SW curve we follow the procedure developed in \cite{Landsteiner:1997vd}. Due to the orientifold, the curve is invariant under $x \to -x$. Also, the charge of the orientifold affects the asymptotic behavior of the NS5-branes, which changes the powers of $v$ compared to the cases without an orientifold. In the end, the SW curve for the quiver in Figure \ref{OrtoQuiver} is given by
\begin{equation}\label{SWSOSp1}
\begin{split}
z^4 + (a_2x^4 + a_1x^2 + a_0)z^3 &+ (b_4x^8 + b_3x^6 + b_2x^4 + b_1x^2 + b_0)z^2 \\
&+ \prod_{i=1}^4(x^2-m_i^2)(c_2x^4 + c_1x^2 + c_0)z + \prod_{i=1}^4(x^2 - m_i^2)^2 = 0\,,
\end{split}
\end{equation}
where $a_0$ and $c_0$ are fixed by the constraint 
\begin{align}
z^4 + a_0z^3 + b_0z^2 + \prod_{i=1}^4m_i^2c_0z + \prod_{i=1}^4m_i^4 = (z-\alpha)^2(z-\beta)^2\,.
\end{align}
The SW differential is still given by \eqref{lambdaSW}. Since its dimension is one, we have
\be
D^{\rm UV}(z) = 4\,, \qquad D^{\rm UV}(x) = 1\,.
\ee
Hence the dimension of the various parameters are 
\begin{equation}
\begin{aligned}
D^{\rm UV}(a_2) &= 0\,, \quad D^{\rm UV}(a_1) = 2\,, \quad D^{\rm UV}(b_4) = 0\,, \nonumber\\  
D^{\rm UV}(b_i) &= 8- 2i \;\; (i=0, 1, 2, 3)\,, \quad D^{\rm UV}(c_2) = 0\,,\quad D^{\rm UV}(c_1) = 2\,.
\end{aligned}
\end{equation}
$a_1$ is the CB operator of the first $USp(2)$, $b_i, (i=0, \cdots, 3)$ are the CB operators of the $SO(8)$ and $c_1$ is the CB operator of the second $USp(2)$. Note that the highest Casimir of $SO(8)$ is reducible, $b_0 = \tilde{b}_0^2$ and $m_i, (i=1, \cdots, 4)$ are the mass parameters for the four flavors\footnote{The field that is charged under both $USp(2)$ and $SO(8)$ is a half-hypermultiplet in the bifundamental representation and it has no mass term.}. Finally, $a_2, b_4, c_2$ are the gauge coupling constants associated to the three CB operators of dimension $2$.  Recall that, in an $\cN=2$ theory, each CB operator with scaling dimension $1 < D \leq 2$ has a corresponding coupling constant with scaling dimension $2-D$.

Let us now deform the above theory as usual by a coupling of the form \eqref{MSdef}, and let us consider turning on a vev for $M$ corresponding to the maximal nilpotent orbit of $USp(8)$, i.e.
\bea
\rho(\sigma^+)=\left(
\begin{array}{cccccccc}
0&1&0&0&0&0&0&0\\
0&0&1&0&0&0&0&0\\
0&0&0&1&0&0&0&0\\
0&0&0&0&0&0&0&1\\
0&0&0&0&0&0&0&0\\
0&0&0&0&-1&0&0&0\\
0&0&0&0&0&-1&0&0\\
0&0&0&0&0&0&-1&0
\end{array}
\right)\,.
\eea
The decomposition of the adjoint representation of $USp(8)$ under the embedded $SU(2)$ is 
\be
\text{adj} = V_1 \oplus V_3 \oplus V_5 \oplus V_7\,,
\ee
and the fluctuation of $M$ will depend on the fields corresponding to the lowest components of the spin $1, 3, 5, 7$ representations. This means that we have:
\bea
M=\left(
\begin{array}{cccccccc}
0&1&0&0&0&0&0&0\\
-7M_{1,-1}&0&1&0&0&0&0&0\\
0&-12M_{1, -1}&0&1&0&0&0&0\\
-7M_{3, -3}&0&-15M_{1, -1}&0&0&0&0&1\\
-M_{7,-7}&0&-7M_{5,-5}&0&0&7M_{1, -1}&0&7M_{3, -3}\\
0&12M_{5,-5}&0&-16M_{3, -3}&-1&0&12M_{1, -1}&0\\
-7M_{5,-5}&0&20M_{3, -3}&0&0&-1&0&15M_{1, -1}\\
0&-16M_{3, -3}&0&-16M_{1, -1}&0&0&-1&0
\end{array}
\right).\nn
\eea

The characteristic polynomial of the above matrix is:
\begin{equation}
\begin{split}
P(x)=&\det\left(x \mathbb{1}_6- M\right) \\
=&\; x^8 + 84M_{1,-1}x^6 + (66M_{3,-3}+1974M_{1,-1}^2)x^4+\\  &+ (-26M_{5,-5}+1364M_{3,-3}M_{1,-1} + 12916M_{1,-1}^3)x^2+\\ &
+ (-M_{7,-7} + 49M_{3,-3}^2 - 98M_{5,-5}M_{1,-1} + 2450M_{3,-3}M_{1,-1}^2 + 11015M_{1,-1}^4)\,,
\end{split}
\end{equation}
and therefore, the SW curve \eqref{SWSOSp1} is deformed as: 
\begin{equation}\label{SWSOSp1d}
\begin{split}
z^4 + (a_2x^4 + a_1x^2 + a_0)z^3 + (b_4x^8 + b_3x^6 + b_2x^4 + b_1x^2 + b_0)z^2 +\\  + P(x)(c_2x^4 + c_1x^2 + c_0)z + P(x)^2 = 0\,.
\end{split}
\end{equation}
As usual, along the RG flow the functional form of the SW differential will change, but those of the six holomorphic ($1,0$)-forms of the curve will remain the same. The CB operator with the largest scaling dimension in the UV is $b_1$, and, given that relative scalings are RG-flow invariant, the candidate field to play the r\^ole of $b_1$ in the IR is the singlet with the largest spin, i.e.~$M_{7,-7}$. Hence, we are led to impose
\begin{align}\label{SOSpansatz}
\frac{{\rm d} \lambda}{{\rm d} b_1} = \frac{ {\rm d}\lambda'}{{\rm d} M_{7,-7}}\,,
\end{align}
where $\lambda'$ is the SW differential in the IR. Since the SW differential in the UV \eqref{lambdaSW} does not have an explicit dependence on $b_1$, but depends on it only through $v$, we can write the l.h.s.~of \eqref{SOSpansatz} as 
\begin{align}
\frac{{\rm d} \lambda}{{\rm d} b_1}  = -\frac{x^2 z dz}{16x^{15} + \cdots}\,.
\end{align}
Since the scaling dimension of $\lambda'$ is $1$, we obtain
\begin{align}\label{SOSprelation1}
1 - D^{\rm IR}(M_{7,-7}) = 2D^{\rm IR}(z) - 13D^{\rm IR}(x)\,. 
\end{align}
The explicit form of the curve \eqref{SWSOSp1d} implies
\begin{align}\label{SOSprelation2}
D(z) = 4D(x)\,, \qquad 8D(x) = D(M_{7,-7})\,.
\end{align}
Combining \eqref{SOSprelation1} with \eqref{SOSprelation2} yields
\begin{align}
D^{\rm IR}(z) = \frac{4}{3}\,, \qquad D^{\rm IR}(x) = \frac{1}{3}\,.
\end{align}
Then the scaling dimension of the various parameters in the SW curve \eqref{SWSOSp1d} is given by 
\bea
&&D^{\rm IR}(a_2) = 0\,, \quad D^{\rm IR}(a_1) = \frac{2}{3}, \nn\\
&&D^{\rm IR}(b_4) = 0\,, \quad D^{\rm IR}(b_3) = \frac{2}{3}\,, \quad D^{\rm IR}(b_2) = \frac{4}{3}\,, \quad D^{\rm IR}(b_1) = 2\,, \quad D^{\rm IR}(\tilde{b}_0) = \frac{4}{3}\,,\nn\\
&&D^{\rm IR}(c_2) = 0\,, \quad D^{\rm IR}(c_1) = \frac{2}{3}\,, \nn\\
&&D^{\rm IR}(M_{1,-1}) = \frac{2}{3}\,, \quad D^{\rm IR}(M_{3,-3}) = \frac{4}{3}\,, \quad D^{\rm IR}(M_{5,-5}) = 2\,, \quad D^{\rm IR}(M_{7,-7}) = \frac{8}{3}\,.\nn\\
\eea
From this we conclude that there are still $6$ operators above the unitarity bound, playing the r\^ole of the would-be CB operators in the IR, and hence also in the IR the genus of the SW curve matches the dimension of the base over which it is fibered. However, as one can see from the above scaling dimensions, there is na\"ively no matching between CB operators with $1<d\leq2$ and coupling constants of dimension $2-d$: The are $2$ CB operators of dimension $2$, but $3$ coupling constants of dimension $0$, and also $3$ CB operators of dimension $4/3$, but $4$ coupling constants of dimension $2/3$. To confirm that this expectation is correct, we should make sure that there exists no change of variables leaving the IR SW differential invariant (up to total derivatives), which eliminates from \eqref{SWSOSp1} the two extra coupling constants preventing the match. Unfortunately, this is very hard here, because we do not know the explicit form of the IR SW differential. Nevertheless, we can give some evidence in this direction. First, as opposed to the previously-discussed examples, here we cannot exclude an explicit dependence of $\lambda'$ from the new would-be CB operators. Thus, focusing solely on the operator of largest dimension, and writing with no loss of generality
\be\label{lambdaExpl}
\lambda'=f[x(M_7,, z),z, M_7]\,{\rm d}z\,,
\ee 
Eq. \eqref{SOSpansatz} reads\footnote{Note that $f$ must satisfy five more partial differential equations, which originate from the other holomorphic ($1,0$)-forms. Given the degeneration in dimension of the other would-be CB operators, we do not know the explicit expression of these extra equations.}
\be\label{DiffEqM7}
\frac{\partial f}{\partial x}\left(2 P(x)+(c_2x^4+c_1x^2+c_0)z\right)-\frac{\partial f}{\partial M_7}(16x^{15}+\cdots)=x^2z\,,
\ee
where we have renamed $M_7:=M_{7,-7}$.
Consider changing $x\to x+g(z,M_7)$, leaving everything else invariant. This change of variable induces a modification of \eqref{lambdaExpl} which amounts to a total derivative if and only if $f$ depends linearly on $x$. But such a dependence can never satisfy Eq. \eqref{DiffEqM7} for generic values of the parameters. A similar argument can be drawn swapping $x$ and $z$. However, one can think of a more general change of variables, such as
\be\label{ChangeGen}
\begin{split}
x\to x+g_x(M_7,z)\,,\\
z\to z+g_z(M_7,x)\,.
\end{split}
\ee
The change of the IR SW differential then reads
\be
\Delta\lambda'_{SW}=\left[f(x+g_x,z+g_z,M_7)-f(x,z,M_7))\right]{\rm d}z+f(x+g_x,z+g_z,M_7)\frac{\partial g_z(x,M_7)}{\partial x}\,{\rm d}x\,.
\ee
One condition for the above to be a total derivative is that the change of variables \eqref{ChangeGen} must be such that $f(x+g_x(z,M_7),z+g_z(x,M_7),M_7)$ looses any explicit dependence on $z$. Though we lack a proof of this, we argue that this cannot happen compatibly with the six differential equations that the function $f$ must satisfy.

Consequently, the mismatch between CB operators and couplings that we found in the IR would explain why this theory does not exhibit supersymmetry enhancement, despite we found the right CB dimension. 

The matching condition refines our necessary criterion for enhancement, and seems to give a geometric meaning to the condition (3.12) of \cite{Giacomelli:2018ziv}, which every theory displaying supersymmetry enhancement should meet.

\section{Conclusions}
\label{s:Conclusions}

In this paper we have extended our geometric understanding of the phenomenon of SUSY enhancement to 4d field theories of rank higher than $1$. In \cite{Carta:2018qke} the origin of the enhancement for rank-$1$ theories was traced in the holonomy reduction of the F-theory internal space used to engineer the field theory. Here, instead, we have used class-$\cS$ constructions to track the enhancement down to a hyperk\"ahler-structure restoration on the moduli space of solutions of the underlying Hitchin system. As in \cite{Carta:2018qke}, we have formulated a simple necessary algebraic criterion for enhancement in terms of an auxiliary geometry given by a Riemann-surface fibration: If SUSY enhancement occurs in the IR, this geometry needs to factorize in such a way that the dimension of the base of the fibration reduces and becomes equal to the genus of the fiber. We have refined this criterion, supplementing it by a matching condition between CB operators of dimension $1<D\leq2$ and coupling constants of dimension $2-D$\footnote{Such a condition is trivially satisfied for all theories of rank $1$ which exhibit enhancement.}. For theories exhibiting enhancement, we have been able to write down the complete SW geometry (including masses and couplings) of the IR theory, and compute all conformal dimensions of CB operators by purely algebraic techniques, i.e.~without relying on any maximization procedure.

An important remark is in order. The above-mentioned factorization implies that some of the fields hitting the unitarity bound disappear from the IR theory. Our technique is able in a purely geometric manner to distinguish them from those becoming instead masses and coupling constants of the IR theory: The functional form of the SW differential gets modified by the flow in such a way that new changes of coordinates become available in the IR and this allows us to get rid precisely of those monomials containing the decoupled fields. It would be very interesting to further investigate the deeper geometric meaning of these specific RG-flow-induced modifications of the SW differential. We hope to come back to this matter in a future publication.

As already mentioned in the introduction, our treatment of class-$\cS$ theories in this paper does not cover irregular punctures featuring a nontrivial degeneracy among the eigenvalues of the Hitchin field (Type III irregular punctures \cite{Xie:2012hs}). Let us briefly illustrate here what the issue is. Consider a Hitchin field on the sphere (parametrized by the coordinate $z$) with a puncture of Type III at $z=\infty$ (we consider the $SU(N)$ case for simplicity). Locally around the puncture the field can be diagonalized and expanded in powers of $z$ as follows:
\be\label{hitfd}
\Phi=M_{n+1}z^ndz+\dots + M_0\frac{dz}{z}+ M'\frac{dz}{z^2}+\dots\,,
\ee
where $n>0$, and the $M_i$'s and $M'$ are diagonal $N\times N$ traceless matrices. The matrices $M_0,\dots M_{n+1}$ encode the data defining the boundary condition at the irregular puncture. The matrix $M'$ and subsequent terms are determined instead by solving the differential equation of the Hitchin system; the corresponding terms are not singular at infinity. In the case of type III punctures, the matrices $M_0,\dots M_{n+1}$ have degenerate eigenvalues and the degeneracy for the matrix $M_i$ is \emph{not} arbitrary, but instead it is constrained by the degeneracies of $M_{i+1}$. 

Assuming the UV theory has a puncture of Type III, upon activating the Maruyoshi-Song RG flow, we are left with a twisted Hitchin field $\Phi_1$ with the same boundary condition as in (\ref{hitfd}) at $z=\infty$: 
\be\label{hittw}
\Phi_1=M_{n+1}z^{n+1}\sqrt{dz}+\dots+ M_0\sqrt{dz}+\frac{M'}{z}\sqrt{dz}+\dots\,.
\ee 
Again, the term proportional to $M'$ is not singular at infinity and the matrix $M'$ is not part of the data defining the boundary condition, it is determined by solving the differential equation of the twisted Hitchin system. 

On the one hand, according to our prescription, the new Hitchin field $\widetilde{\Phi}$ in the infrared should read 
\be
\widetilde{\Phi}=M_{n+1}z^{n+1}dz+\dots+M_0dz+M'\frac{dz}{z}+\dots\,. 
\ee 
On the other hand, the term proportional to $M'$ is now singular at infinity and is therefore part of the data defining the boundary condition. This in particular means that the eigenvalue degeneracy for $M'$ cannot be arbitrary and is actually constrained by the form of the matrix $M_0$. For the twisted and ordinary Hitchin systems to be equivalent, it must be the case that the form of $M'$ in (\ref{hittw}) as determined by the differential equation of the twisted Hitchin system is automatically consistent with the constraint imposed by $M_0$. This is not necessarily true. We conclude that in the case of Type III punctures the twisted Hitchin system and the ordinary one are generically inequivalent, thus preventing any enhancement. However, we are unable to decide whether accidental equivalences may occur, leading to SUSY enhancements for theories with type III punctures. We plan to come back to this issue in the near future.

\bigskip

\centerline{\bf \large Acknowledgments}

We would like to thank A. Collinucci for initial collaboration and many useful discussions. We also thank P. Agarwal, I. Bah, F. Bonetti, H-C. Kim, K. Maruyoshi, J. Song, W. Yan, and M. Zimet for discussions.

HH and RS are grateful to the Aspen Center for Physics (which is supported by National Science Foundation grant PHY-1607611), for hospitality during the initial stage of this work. We are pleased to acknowledge the 2019 Pollica summer workshop where some of this work was performed, and we are grateful to its supporting organizations: the Simons Foundation (Simons Collaboration on the Nonperturbative Bootstrap) and the INFN. The work of FC is supported by the ERC Consolidator Grant STRINGFLATION under the HORIZON 2020 grant agreement no. 647995. The work of SG is supported by the ERC Consolidator Grant 682608 ``Higgs bundles: Supersymmetric Gauge Theories and Geometry (HIGGSBNDL)''. The work of HH is supported in part by JSPS KAKENHI Grant Number JP18K13543. The work of RS is supported by the program ``Rita Levi Montalcini'' for young  researchers (D.M. n. 975, 29/12/2014).

\newpage

\bibliography{Refs.bib}
\bibliographystyle{JHEP}

\end{document}